\documentclass[12pt]{article}

\usepackage[english]{babel}
\usepackage[T1]{fontenc} 
\usepackage{lmodern}
\usepackage[cp1251]{inputenc}  
\usepackage{mathtools}
\usepackage{amsfonts,amssymb,amsmath} 
\usepackage{graphicx} 
\usepackage{cite}
\usepackage{hyperref}
\hypersetup{colorlinks=true, linkcolor=blue, filecolor=blue, urlcolor=blue} 

\newtheorem{theorem}{Theorem} 

\allowdisplaybreaks

\usepackage{diagbox}

\begin{document}

\title{Optimal State Manipulation for a~Two-Qubit System Driven by Coherent and Incoherent Controls}

\author{ Oleg~V.~Morzhin$^{1,2,}$\footnote{E-mail: \url{morzhin.oleg@yandex.ru};~ 
        \href{http://www.mathnet.ru/eng/person30382}{mathnet.ru/eng/person30382};~ 
        \href{https://orcid.org/0000-0002-9890-1303}{ORCID 0000-0002-9890-1303}} 
        \quad and \quad 
	Alexander~N.~Pechen$^{1,2,}$\footnote{E-mail: \url{apechen@gmail.com};~ 
        \href{http://www.mathnet.ru/eng/person17991}{mathnet.ru/eng/person17991};~ 
        \href{https://orcid.org/0000-0001-8290-8300}{ORCID 0000-0001-8290-8300}} 
        \vspace{0.2cm} \\
	$^1$ Department of Mathematical Methods for Quantum Technologies,\\
        Steklov Mathematical Institute of Russian Academy of Sciences, \\
	8 Gubkina St., Moscow, 119991, Russia;\\
	$^2$ Quantum Engineering Research and Education Center, \\
        University of Science and Technology MISIS,\\
	6 Leninskiy Prospekt, Moscow, 119991, Russia}

\date{}
\maketitle

\begin{abstract}
Optimal control of two-qubit quantum systems attracts high interest 
due to applications ranging from two-qubit gate generation to optimization 
of receiver for transferring coherence matrices 
along spin chains. State preparation and manipulation is among important tasks 
to study for such systems. Typically coherent control, e.g. a shaped laser pulse, is used to manipulate two-qubit systems. However, the environment can also 
be used --- as an incoherent control resource. In this article, we consider 
optimal state manipulation for a two-qubit system whose dynamics is governed 
by the  Gorini--Kossakowski--Sudarshan--Lindblad master equation, where coherent 
control enters into the Hamiltonian and incoherent control into both 
the Hamiltonian (via Lamb shift) and the superoperator of dissipation. We exploit
two physically different classes of interaction with coherent control and optimize 
the Hilbert--Schmidt overlap between final and target density matrices, including optimization of its steering to a given value. We find the conditions when zero coherent and incoherent controls satisfy the Pontryagin maximum principle, and in addition, when they form a~stationary point of the objective 
functional. Moreover, we find a case when this stationary point provides the globally minimal value of the overlap. Using upper and lower bounds 
for the overlap, we develop one- and two-step gradient projection methods 
operating with functional controls. 

\vspace{0.3cm}

\end{abstract}

\section{Introduction} 
\label{Introduction}

Developing methods of quantum control 
is crucial for modern quantum technologies~\cite{KochEPJQuantumTechnol2022, 
BrifNewJPhys2010, 
ButkovskiyBook1990, TannorBook2007, 
FradkovBook2007, WisemanBook2009, BonnardBook2012, KochJPhysCondensMatter2016, 
DAlessandroBook2021, KuprovBook2023}. In particular, quantum control is used for fast gate generation and modeling basic 
operations for quantum computing~\cite{PalaoPRL2002, 
TreutleinPRA2006, Grace2007,WuPRA2008, PoulsenPRA2010, PechenJPA2017,GoogleNature, BasilewitschNewJPhys2019, 
RiazQuantumInfProcess2019, VolkovJPA2021, Volkov2022}. 
Various control problems for two-qubit and four-level open quantum systems 
were considered, e.g., in~\cite{LiScienceChina2013, 
RafieePRA2016, AllenPRA2017, HuIJTP2018, FengPRA2018, SunQuantumInfProcess2020, ShindiIEEE2022,
Hirose2018, MorzhinLJM2021, PetruhanovIntJModPhysB2022}. Optimal control theory was used to derive an optimization algorithm that determines the best entangling two-qubit gate for a given physical model~\cite{PRA84_042315}. Transfer of the two-qubit quantum information 
in spin chain with the length~$N$ was studied~\cite{WangIndianJPhys2022}. Controllability 
of special four-level closed quantum systems was studied for systems with degenerate transitions~\cite{SchirmerPRA2001,PolackPRA2009} or with degenerate energy levels~\cite{KuznetsovLJM2022,MyachkovaFourLevel}. Optimization of few qubit systems was used for transferring coherence matrices over spin chains~\cite{BochkinQIP2022}. System--field couplings were exploited to elucidate electronic quantum coherence effects in two-qubit systems for the model of photosynthesis~\cite{Ai2014}. Various optimal  
control  methods were adapted for quantum control problems including the Pontryagin maximum principle~(PMP)~\cite{ButkovskiyBook1990,BoscainPRXQuantum2021,BuldaevMathematics2022}, 
gradient flows~\cite{SchulteHerbruggenRevMathPhys2010}), Hamilton--Jacobi--Bellman equation~\cite{Gough2005},
Krotov method~\cite{TannorBook2007,KrotovBook1996} (more references in the survey~\cite{MorzhinUMN2019}), etc. Other systems, such as three-level, are also of special interest~\cite{Song2016},~etc.  For open quantum systems, a general approach based on gradient optimization over complex Stiefel manifolds was developed for solving problems which appear in quantum technologies, for which explicit analytical expressions for gradient and Hessian of quantum control objectives and the corresponding optimization techniques were developed and various examples were studied including cases with constraints~\cite{Pechen_Prokhorenko_Wu_Rabitz_2008,Oza_Pechen_Dominy_Beltrani_Moore_Rabitz_2009}.

Closed quantum systems can be driven by coherent control, e.g. by shaped laser light. 
In open quantum systems, coherent control can appear in the Hamiltonian~\cite{KochJPhysCondensMatter2016} or in both Hamiltonian and dissipative part of the master equation~\cite{DannPRA2020}.  Open quantum systems, in addition to coherent control, can be driven by various forms of incoherent control which can be realized, e.g., by tailored environment as proposed in~\cite{PechenPRA062102.2006, PechenPRA2011}. Another approach is to use back-action of non-selective quantum measurements, either alone or in combination with coherent control~\cite{PechenPRA052102.2006, ShuangJChemPhys2007, 
Shuang_Zhou_Pechen_Wu_Shir_Rabitz_2008}. Various forms of incoherent control and engineered reservoirs were used for superabsorption of light~\cite{HigginsNatComm2014}, control of non-Markovian~\cite{HwangPRA2012,NJP17_063031} and Markovian  open quantum systems~\cite{LucasRPL2013}, control of dissipation in cavity QED~\cite{LiningtonPRA2008}, incoherent control in a Bose-Hubbard dimer~\cite{ZhongPRA2011}, photoionization of atoms under noise~\cite{SinghPRA2007}, generating quantum coherence through an autonomous thermodynamic machine~\cite{MukhopadhyayPRA2018}, studying decoherence~\cite{JPA2007_8033}, controlling optical signals with quantum heat-engine~\cite{QutubuddinPRR2021}, suppression of decoherence via quantum Zeno effect~\cite{Facchi2005}, optimization of up-conversion hues in phosphor~\cite{LaforgeJCP2018}, Landau-Zener transitions~\cite{Pechen_Trushechkin_2015}, environment-assisted quantum walks were applied for photosynthetic energy transfer~\cite{Mohseni2008}, etc. An approximate controllability for generic $N$-level quantum systems driven
by coherent and incoherent controls was established~\cite{PechenPRA2011}. The two-level case was rigorously studied recently~\cite{LokutsievskiyJPA2021}. Various control problems for coherent and incoherent control of one-qubit open quantum systems were studied, e.g.
in~\cite{LokutsievskiyJPA2021, MorzhinLJM2019, MorzhinPhysPartNucl2020, MorzhinIzvRAN2023}. The Gradient Ascent Pulse Engineering (GRAPE) approach for two-qubit quantum systems with time dependent coherent and incoherent
controls was developed~\cite{PetruhanovIntJModPhysB2022}.

In this work, we consider two-qubit systems driven by incoherent control which, in general, is piecewise continuous function of time. The control objectives correspond to maximizing or minimizing the Hilbert--Schmidt overlap between the final density matrix and a given target density matrix, or steering the overlap as close as possible to a given admissible value. For the problems of maximizing 
the overlap, we derive the conditions for the target density matrix under which zero coherent and incoherent controls satisfy the PMP and, in addition, are singular. For these problems, we derive the gradients of the objective functionals and develop the one-step gradient projection method (GPM-1) using the works  
\cite{LevitinUSSRComputMathMathPhys1966, DemyanovBook1970, PolyakBook1987} and the two-step GPM (GPM-2) using the works  \cite{PolyakBook1987, PolyakUSSRComputMathMathPhys1964,  AntipinDifferEqu1994, HeutsIEEE2021} and that the heavy-ball method, as known, can significantly accelerate the one-step gradient optimization by using the results of the two last iterations for making the next update.

The structure of this work is the following. In Sect.~\ref{Sec:IC}, we overview the incoherent control method. Sect.~\ref{Section3} contains 
description of the model and formulation of the control problems. In Sect.~\ref{Section4}, we formulate the PMP for these problems (Theorem~\ref{Theorem1}),
analytically solve the system for zero coherent and incoherent controls 
(Subsect.~\ref{Subsection4.2}) and, for the problems of optimizing the overlap, derive conditions when
zero coherent and incoherent controls satisfy the PMP (Theorem~\ref{Theorem2}), derive conditions when these controls form a~stationary point (Theorem~\ref{Theorem3}), 
and give an analytical example when these singular controls are exactly optimal (Subsect.~\ref{Subsection4.6}). Sect.~\ref{Section5} describes  GPM-1 and GPM-2   for various objective criteria. The numerical results are provided in~Sect.~\ref{Section6}. The Conclusion Sect.~\ref{Section7_Conclusions} summarizes the work.

\section{Incoherent Control by the Environment}\label{Sec:IC}

As mentioned above, various forms of using the environment for control were considered by many researchers.  In this work, we use the incoherent control method which was proposed in~\cite{PechenPRA062102.2006}. As it is known in the theory of open quantum systems, master equation for the reduced density matrix of a quantum system interacting with the environment is determined by (1) the system Hamiltonian, (2) the interaction Hamiltonian between the system and the environment and (3) state of the environment. In general, the system-bath interaction is fixed once the system and bath are prepared. However, state of the environment can be not fixed and used for control. Therefore even if the  interaction Hamiltonian between the system and the environment  is fixed, one can vary the coefficients of the master equation by changing the state of the environment. This is the key idea of incoherent control.

\begin{figure}[b!]
\centering
\includegraphics[width = 0.5\linewidth]{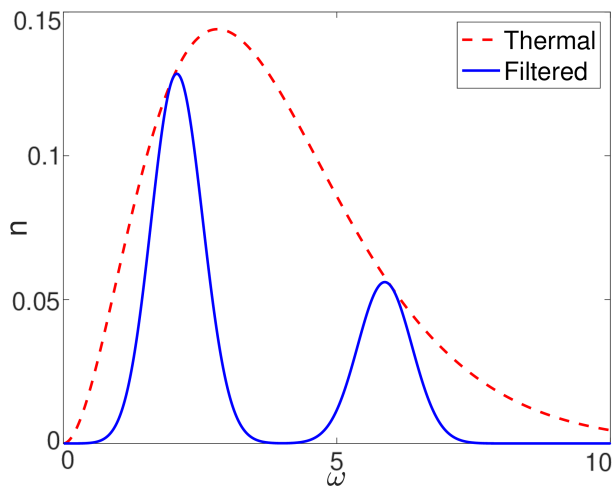}
\caption{Planck density of black-body radiation for $\beta = 1$ and its Gaussian filtering with $f(\omega)=\exp\bigl(-\frac{(\omega-\omega_1)^2}{2\sigma^2}\bigr)+\exp\bigl(-\frac{(\omega-\omega_2)^2}{2\sigma^2}\bigr)$, where $\omega_1=2$, $\omega_2 = 6$ and variance $\sigma^2=0.25$. }
\label{Fig1}
\end{figure}

The natural environment is formed by physical particles surrounding the controlled system. Depending on the particular situation, they can be photons, phonons, surrounding gas, spins, etc. Typically state of the environment is considered as thermal.  Time-independent frequency distribution of thermal photons with inverse temperature  $\beta$ has the famous Planck form, which in the Planck system of units which we use in this work (where reduced Planck constant $\hbar$, speed of light $c$, and~Boltzmann constant $k_\textrm{B}$ equal one), is
$n_\omega= \dfrac{1}{\pi^2}\dfrac{\omega ^3}{\exp{(\beta \omega)} - 1}$. But thermal form is a very limiting  assumption and more general non-equilibrium forms of the state of the environment can be used.  If temperature is varied, so that $\beta=\beta(t)$, then spectral density becomes time-dependent,
\[
n_\omega(t)= \frac{1}{\pi^2}\dfrac{\omega ^3}{\exp{(\beta (t) \omega)} - 1}.
\]
Variation of temperature is only the simplest option to vary $n_\omega(t)$ and generally spectral density can be made non-equilibrium and time-dependent  with some other shapes $n_\omega(t)$. Such non-thermal shapes can be obtained, e.g.,~by filtering black-body radiation, using light-emission diodes, etc, see Fig.~\ref{Fig1}. Of course, not always state of the environment can be easily variable experimentally. It can be relatively easy done for photons, more difficult for phonons, and perhaps more difficult for a quantum gas. Experimental realization in any particular case is important and beyond the scope of our consideration. 

This incoherent control approach can be applied to various system-environment models. For example, in~\cite{PechenPRA062102.2006} it was applied to master equations with dissipators of the forms which were derived in the WCL and LDL limits known in theory of open quantum systems. However, incoherent control method is not limited to these two forms of the master equation and has a general character --- it is any control when state of the incoherent environment surrounding the system is used as a control resource.

Physically incoherent control can be implemented by using time-dependent spectral density of incoherent photons or phonons $n_\omega(t)$. This density is considered as a function of both transition frequency $\omega$ and time $t$. Of course, not for all transition frequencies this spectral density can  easily be experimentally made arbitrary, so practical applicability depends on the exact type of the environment (photons or phonons), spectrum of the system, and experimentally available techniques to generate various $n_\omega$. For photons, their spectral density generally can also include dependence on polarization $\alpha$, so that incoherent control can be a function of $\omega$, $\alpha$, and~$t$. Here, we do not use this option. 

Since incoherent control has physical meaning of density of particles of the environment, mathematically it should be a non-negative quantity. In~addition, one can require that  total density of particles of the environment at any time moment is finite, so that
\[
n_\omega(t)\ge 0,\qquad \int\limits_0^\infty n_\omega(t)d\omega<\infty\qquad \forall t \geq 0.
\]

Incoherent control allows to make decoherence rates in the master equation time-dependent (see page~3 in~\cite{PechenPRA062102.2006}) as
\begin{align}\label{Eq:ME2}
\dot{\rho_t} = -i [H_c(t), \rho_t] + \sum\gamma_{ij}(t) {\cal D}_{ij}(\rho_t).
\end{align}
The time-dependent decoherence rates for transition between levels $i$ and $j$ with transition frequency $\omega_{ij}$ are expressed via $n(t)$, so that
\[
\gamma_{ij}(t)=\pi\int d{\bf k}\delta(\omega_{ij}-\omega_{\bf k}) \lvert g({\bf k}) \rvert^2(n_{\omega_{ij}}(t)+\kappa_{ij})
\]
where $\kappa_{ij}=1$ for $i>j$ and $\kappa_{ij}=0$ otherwise, $\omega_{\bf k}$ is the dispersion law for the bath (e.g., $\omega = \lvert {\bf k} \rvert c$ for photons, where ${\bf k}$ is photon momentum), and $g({\bf k})$ describes coupling of the system to $\bf k$-th mode of the bath. The magnitude of the decoherence rates affects the speed of decay of off-diagonal elements of the system density matrix and determines the values of the diagonal elements towards which the diagonal part of the system density matrix evolves. It can be used, for~example, for~approximate generation of various density matrices~\cite{PechenPRA2011}.

\section{Control System and Objective Functionals}
\label{Section3}

We study the system of two-qubits evolving under coherent and incoherent controls~\cite{MorzhinLJM2021,PetruhanovIntJModPhysB2022}. We consider both coherent and incoherent controls
as piecewise continuous functions. Let $\mathcal{H} = \mathbb{C}^2\otimes \mathbb{C}^2$ be the Hilbert space
of the two-qubit system.  Density matrix $\rho(t)$ is a $4\times 4$ positive semi-definite matrix 
with unit trace, $\rho(t) \in \mathbb{C}^{4 \times 4}$, 
$\rho(t) \geq 0$, ${\rm Tr}\rho(t) = 1$. System's dynamics is governed by the master equation of the form~(\ref{Eq:ME2})
\begin{align}
	\dot \rho(t) = 
	-i \left[H_S + \varepsilon H_{{\rm eff},n(t)} + H_{u(t)}, \rho(t) \right] + 
	\varepsilon \mathcal{L}_{n(t)}(\rho(t)), \qquad \rho(t) = \rho_0.
	\label{GKSL_eq}
\end{align}
Here $H_S$ is the free two-qubit Hamiltonian, $H_{{\rm eff},n(t)}$ is the two-qubit effective Hamiltonian (Lamb shift), 
which is determined by incoherent control $n = \left( n_{\omega_1}, n_{\omega_2} \right)$, 
$H_{u(t)} = V u(t)$ is coupling to coherent control~$u$ via interaction operator $V$,
$\mathcal{L}_{n(t)}(\rho(t))$ is the superoperator of dissipation depending on~$n$, $\varepsilon > 0$ describes strength of the coupling between 
the system and its environment, $\rho_0$ is the initial state which can be either pure or mixed. We denote the complete control $c = (u, n_1, n_2)$.  The notation $[A, B] 
= AB - BA$ denotes the commutator of matrices $A$ and $B$.

The free and effective Hamiltonians are:
\begin{align}
	H_S &= H_{S,1} + H_{S,2} = \frac{\omega_1}{2} \left( \sigma^z \otimes \mathbb{I}_2 \right) 
	+ \frac{\omega_2}{2} \left( \mathbb{I}_2 \otimes \sigma^z \right),
	\label{H_S} \\
	H_{{\rm eff},n(t)} &= \sum\limits_{i = 1}^2 H_{{\rm eff},n(t),i} = \Lambda_1 n_{\omega_1}(t) \left( \sigma^z \otimes \mathbb{I}_2 \right) + 
	\Lambda_2 n_{\omega_2}(t) \left( \mathbb{I}_2 \otimes \sigma^z \right)
	\label{H_eff}
\end{align}
where $\sigma^z = \begin{pmatrix}
	1 & 0 \\
	0 & -1
\end{pmatrix}$
is Z Pauli matrix, $\mathbb{I}_2$ is the $2\times 2$ identity matrix, 
$\sigma^z \otimes \mathbb{I}_2 = 
\begin{pmatrix}
	\mathbb{I}_2 & 0_2 \\
	0_2 & -\mathbb{I}_2
\end{pmatrix}$,
$\mathbb{I}_2 \otimes  \sigma^z = 
\begin{pmatrix}
	\sigma^z & 0_2 \\
	0_2 & \sigma^z
\end{pmatrix}$, $0_2$ is the $2\times 2$ zero matrix. Incoherent controls 
$n_1 = n_{\omega_1}$ and $n_2 = n_{\omega_2}$ represent density of particles 
of the environment at frequencies $\omega_1$ and $\omega_2$ and can be adjusted 
independently. We denote $n = (n_1, n_2)$. By the physical meaning, one has
\begin{align}
	n_1(t) \geq 0, \qquad n_2(t) \geq 0 \qquad \text{for all} \quad t \in [0, T].
	\label{lower_constraints_to_incoherent_controls}
\end{align}

The interaction operator $H_{u(t)} = V u(t)$ is considered in general with arbitrary Hermitian matrix~$V$ and in particular, as in~\cite{MorzhinLJM2021}, 
we consider the following two versions:
\begin{align}  
	V=V_1= \sigma^x \otimes \mathbb{I}_2 + \mathbb{I}_2 \otimes \sigma^x =  
	\begin{pmatrix}
		\sigma^x & \mathbb{I}_2 \\
		\mathbb{I}_2 & \sigma^x
	\end{pmatrix}, \qquad  
	V=V_2= \sigma^x \otimes \sigma^x = \begin{pmatrix}
		0_2 & \sigma^x \\
		\sigma^x & 0_2
	\end{pmatrix},
	\label{V_variants_1_and_2}
\end{align} 
where $\sigma^x = \begin{pmatrix}
	0 & 1 \\
	1 & 0
\end{pmatrix}$ 
is $X$ Pauli matrix. In the case $V=V_1$, the same coherent control~$u$ addresses each 
qubit independently, while in the case $V=V_2$, the control~$u$ acts to couple the qubits via XX coupling. 

The superoperator of dissipation acts on density matrix as
\begin{align}
	\mathcal{L}_{n(t)}(\rho(t)) &= 
	\mathcal{L}_{n(t),1}(\rho(t)) + 
	\mathcal{L}_{n(t),2}(\rho(t)), 
	\label{dissipator}
	\\
	\mathcal{L}_{n(t),j}(\rho(t)) &=  \Omega_j (n_{\omega_j}(t) + 1) \left( 2 \sigma^-_j \rho \sigma^+_j - 
	\sigma_j^+ \sigma_j^- \rho - \rho \sigma_j^ + \sigma_j^- \right) \nonumber \\
	& \quad+ \Omega_j n_{\omega_j}(t) \left( 2\sigma^+_j \rho \sigma^-_j - 
	\sigma_j^- \sigma_j^+ \rho - \rho \sigma_j^- \sigma_j^+ \right),
	\quad j = 1,2, 
	\label{dissipator_j}
\end{align}
where $\Lambda_j>0$ and $\Omega_j>0$ are some constants and 
$\sigma_1^{\pm} = \sigma^{\pm} \otimes \mathbb{I}_2$, 
$\sigma_2^{\pm} = \mathbb{I}_2 \otimes \sigma^{\pm}$ 
with $\sigma^+ = \begin{pmatrix}
	0 & 0 \\ 1 & 0
\end{pmatrix}$, 
$\sigma^- = \begin{pmatrix}
	0 & 1 \\ 0 & 0
\end{pmatrix}$. 
The notation $\{A, B\} = AB + BA$ denotes anti-commutator of matrices $A$ and $B$.

The objective which we study is the Hilbert--Schmidt overlap between $\rho(T)$, evolved with some admissible control~$c$,  
and a~given target density matrix~$\rho_{\rm target}$: $F(\rho(T), \rho_{\rm target}) := \langle \rho(T), 
\rho_{\rm target} \rangle = {\rm Tr}(\rho(T) \rho_{\rm target})$.  We consider the problems of maximizing and minimizing the overlap, and steering the overlap to a given value~$M \in (0, 1)$:
\begin{align}
	J(c) & = F(\rho(T); \rho_{\rm target}) = \langle \rho(T), \rho_{\rm target} \rangle \to \sup/\inf,
	\label{J_inf_sup}\\
J_{M,1}(c) &= \left\lvert F(\rho(T); \rho_{\rm target}) - M \right\rvert = \left\lvert \langle \rho(T), \rho_{\rm target} \rangle - M \right\rvert \to \inf,   
\label{J_1_close_to_M} \\
J_{M,2}(c) &= \left( F(\rho(T); \rho_{\rm target}) - M \right)^2 = \left( \langle \rho(T), \rho_{\rm target} \rangle - M \right)^2 \to \inf. 
\label{J_2_close_to_M}
\end{align} 

We use the realification of density matrix as performed in~\cite{MorzhinLJM2021}.
Then the system (\ref{GKSL_eq}) with (\ref{H_S},\ref{H_eff}), (\ref{dissipator},\ref{dissipator_j}),  and with interaction $V=V_1$ or $V=V_2$ is reduced 
to a~system of linear ordinary differential equations for real vector representing real and imaginary 
parts of the matrix elements of density matrix. Hermiticity of density matrix gives that 
\begin{align}
	\rho = \begin{pmatrix}
		\rho_{1,1} & \rho_{1,2} & \rho_{1,3} & \rho_{1,4} \\
		\rho_{1,2}^{\ast} & \rho_{2,2} & \rho_{2,3} & \rho_{2,4} \\
		\rho_{1,3}^{\ast} & \rho_{2,3}^{\ast} & \rho_{3,3} & \rho_{3,4} \\
		\rho_{1,4}^{\ast} & \rho_{2,4}^{\ast} & \rho_{3,4}^{\ast} & \rho_{4,4} 
	\end{pmatrix} 
	= \begin{pmatrix}
		x_1 & x_2 + i x_3 & x_4 + i x_5 & x_6 + i x_7 \\
		x_2 - i x_3 & x_8 & x_9 + i x_{10} & x_{11} + i x_{12} \\
		x_4 - i x_5 & x_9 - i x_{10} & x_{13} & x_{14} + i x_{15} \\
		x_6 - i x_7 & x_{11} - i x_{12} & x_{14} - i x_{15} & x_{16} 
		\label{rho_parametrization}
	\end{pmatrix}.
\end{align}
Here $x_j \in \mathbb{R}$, $j=\overline{1,16}$. The 
condition ${\rm Tr}\rho=1$  implies the linear dependence 
\begin{align}
	x_1+x_8+x_{13}+x_{16}=1. \label{trace_condition}
\end{align}

For the two variants of~$V$, as in~\cite{MorzhinLJM2021} we have the corresponding two bilinear homogeneous dynamical systems with a given initial state:
\begin{align}
	\label{dynamical_system_x_common_form}
	\dot x(t) = \left(A + B_u u(t) + B_{n_1} n_1(t) + B_{n_2} n_2(t) \right) x(t), \qquad x(0) = x_0. 
\end{align}  
Here $16 \times 16$ matrices $A$, $B_u$, $B_{n_1}$, $B_{n_2}$ are found after 
substituting (\ref{rho_parametrization}) in~(\ref{GKSL_eq}); $x_0$~is found 
from a given~$\rho_0$. 

The criterion~\cite{ZhangBook2011} 
(Theorem~7.2) for a Hermitian matrix 
to be positive semidefinite implies that $\rho$ is positive semidefinite if and only 
if the determinant of each principal submatrix of $\rho$ is nonnegative. By construction, solutions of the master equation and of the system (\ref{dynamical_system_x_common_form}) 
satisfy these conditions. 

The problems~(\ref{J_inf_sup}--\ref{J_2_close_to_M}) in terms of~(\ref{rho_parametrization}) are:
\begin{align}
J(c) &= \mathcal{F}(x(T); x_{\rm target}) = \langle x(T), \beta \circ x_{\rm target} \rangle \to \sup /\inf,  
\label{J_inf_sup_param_x} \\
J_{M,1}(c) &= \left\lvert \mathcal{F}(x(T); x_{\rm target}) - M \right\rvert = 
\left\lvert \langle x(T), \beta \circ x_{\rm target} \rangle - M \right\rvert \to \inf, 
\label{J_1_close_to_M_param_x} \\
J_{M,2}(c) &= \left(\mathcal{F}(x(T); x_{\rm target}) - M \right)^2 = \left( \langle x(T), \beta \circ x_{\rm target} \rangle - M \right)^2 \to \inf,~~~
\label{J_2_close_to_M_param_x}
\end{align}
where $\beta =$ (1, 2, 2, 2, 2, 2, 2, 1, 2, 2, 2, 2, 1, 2, 2, 1), and ``$\circ$'' denotes 
the Hadamard product. 

The terminal function $J_{M,1}$ is not continuously differentiable, but it describes exactly the difference between the overlap and~$M$. The subdifferential $\partial(\lvert y - M \rvert)$ is $-1$ for $y<M$, 1 for $y>M$, and $[-1,1]$ for $y=M$. On one hand, a possible way for the nonsmooth objective functional~$J_{M,1}(c)$ is to develop nonsmooth optimization based on the corresponding tools (subgradients, etc.)~\cite{NonsmootOpthBook2019}. On another hand, we can make smoothing  of~(\ref{J_1_close_to_M_param_x}) by introducing the following auxiliary objective functional with some  $0< \theta \ll 1$, in analogy with~\cite[Sec.~8]{MorzhinAiT2009}:
\begin{align}
J_{M,1}^{\theta}(c) = \begin{rcases}
\begin{dcases}
	-\mathcal{F}(x(T); x_{\rm target}) + M, & \\ \qquad \qquad \qquad \quad \text{if} \quad \mathcal{F}(x(T); x_{\rm target}) < M - \theta; \\
	\mathcal{F}(x(T); x_{\rm target}) - M, &\\ \qquad \qquad \qquad \quad \text{if} \quad  \mathcal{F}(x(T); x_{\rm target}) > M + \theta; \\
	\frac{1}{2}\left(\frac{(\mathcal{F}(x(T); x_{\rm target}) - M)^2}{\theta} + \theta \right), &\\ \qquad \qquad \qquad \quad  \text{if} \quad 
	\vert\mathcal{F}(x(T); x_{\rm target}) - M\rvert \leq \theta \\
\end{dcases}
\end{rcases} \to \inf.~~~
\label{J_M_1_theta_inf}
\end{align}
This piecewise defined terminal function is already continuously differentiable.

For the system~(\ref{dynamical_system_x_common_form}) and
the control problems (\ref{J_inf_sup_param_x}), 
(\ref{J_2_close_to_M_param_x}), and 
(\ref{J_M_1_theta_inf}), consider the objective functional $I(c)$ to be minimized, which
is either $\overline{J}-J(c)$ ($\overline{J}$ is some upper bound for the overlap), or $J(c)$, or $J_{M,2}(c)$, or $J_{M,1}^{\theta}(c)$.

In post-optimization analysis, in addition to optimized controls $u,~n_1,~n_2$ and the corresponding solution~$\rho$, one can study 
evolution of $F(\rho(t); \rho_{\rm target})$ with time. By analogy, define the function $F^{\theta}_{M,1}(\rho(t); \rho_{\rm target})$
of~$t$ by taking in~(\ref{J_M_1_theta_inf}) $x(t)$ instead of $x(T)$. For $\rho(t)$, 
consider the von Neumann entropy $S(\rho)$, purity $P(\rho)$, Uhlmann--Jozsa fidelity $UJ(\rho_1; \rho_2)$ \cite{MendoncaPRA2008, MorzhinPhysPartNucl2020},
quantum relative entropy $D(\rho_1; \rho_2)$ (quantum analog of the Kullback--Leibler divergence)
\cite{HolevoBook2019, WildeBook2017}, Petz--R\'{e}nyi relative entropy $D_{\alpha}(\rho_1;\rho_2)$
\cite{WildeBook2017}, where we use $\rho(t)$ instead of $\rho$ and $\rho_1$,
and $\rho_{\rm target}$ instead of~$\rho_2$:
\begin{align}
& S(\rho(t)) = - \mathrm{Tr} \left(\rho(t) \log \rho(t) \right)=-\sum_{\lambda_i(t) \neq 0} \lambda_i(t) \log \lambda_i(t) \in [0, \log \mathrm{dim} \mathcal{H}],
	\label{von_Neumann_entropy_t} \\
& P(\rho(t)) = \mathrm{Tr} \rho^2(t) = \langle \rho(t), \rho(t) \rangle =  
	\sum_{i,j} \lvert\rho_{ij}(t) \rvert^2 \in \left[1/\dim \mathcal{H}, 1 \right],
	\label{purity_t} \\ 
& UJ(\rho(t); \rho_{\rm target}) =
	\left({\rm Tr}\sqrt{\sqrt{\rho(t)} \rho_{\rm target} \sqrt{\rho(t)}} \right)^2 \in [0, 1],
	\label{Uhlmann_Jozsa_fidelity_t} \\ 
& D(\rho(t); \rho_{\rm target}) = {\rm Tr}\left(\rho(t) (\log\rho(t) - \log\rho_{\rm target})\right) \geq 0,
	\label{quantum_relative_entropy_t} \\ 
& D_{\alpha}(\rho(t); \rho_{\rm target}) = 
	\frac{1}{\alpha - 1} \log {\rm Tr}\left(\rho^{\alpha}(t) \rho^{1-\alpha}_{\rm target} \right) \geq 0,
	\, \alpha \in (0,1) \cup (1, \infty),
	\label{Petz_Renyi_relative_entropy_t} 
\end{align} 
where $\lambda_i(t)$ are eigenvalues of $\rho(t)$. In our case the dimension of the Hilbert space is ${\dim \mathcal{H}} = 4$. The von Neumann entropy 
is minimal at pure states where it equals to zero; its maximum value is reached 
at the completely mixed quantum state $\rho = \mathbb{I} / \mathrm{dim} \mathcal{H}$; in our case $\log \mathrm{dim} \mathcal{H} = \log 4 \approx 1.386$. The maximal value~1 
of purity is obtained at any pure state and the minimum value $1 / \mathrm{dim} \mathcal{H}$ 
is obtained at the completely mixed state. One has  
$UJ(\rho_{\rm target}; \rho_{\rm target}) = 1$,  
$D(\rho_{\rm target}; \rho_{\rm target}) = 0$,  
$D_{\alpha}(\rho_{\rm target}; \rho_{\rm target}) = 0$, and
$D(\rho(t); \rho_{\rm target}) = \lim\limits_{\alpha \to 1} D_{\alpha}(\rho(t); \rho_{\rm target})$.

\section{Pontryagin Maximum Principle and Zero Controls}
\label{Section4}

\subsection{Pontryagin Maximum Principle}
\label{Subsection4.1}

The Pontryagin function~\cite{PontryaginBook1962} for our problem is 
$h(p, x, u, n_1, n_2) = \langle p, \left(A + B_u u + B_{n_1} n_1 + B_{n_2} n_2\right) x \rangle = 
\mathcal{K}^{u} u + \mathcal{K}^{n_1} n_1 + \mathcal{K}^{n_2} n_2 + \overline{h}(p,x)$,
where $p, x \in \mathbb{R}^{16}$ and $u, n_1, n_2 \in \mathbb{R}$ (i.e. here
$p, x, u, n_1, n_2$ are real values and not functions), the switching functions are
$\mathcal{K}^{u} = \frac{\partial h}{\partial u} = \langle p, B_u x \rangle$,  
$\mathcal{K}^{n_j} = \frac{\partial h}{\partial n_j} = \langle p, B_{n_j} x \rangle$, $j=1,~2$;  
$\overline{h} = \langle p, A x \rangle$.  
The conjugate system is
\begin{align}
	\dot p(t) = - \left( A^T + B_u^T u(t) + B_{n_1}^T n_1(t) + B_{n_2}^T n_2(t) \right) p(t), 
	\label{conjugate_system_ODE}
\end{align}
where the transversality condition for the problems of maximization ($s=1$) and minimization ($s=-1$) of $J(c)$ is:
\begin{align}
	p(T) = s \nabla \mathcal{F}(x; x_{\rm target})\big\vert_{x=x(T)} = s \beta \circ x_{\rm target}.
	\label{transversality_J_min_or_max}	
\end{align} 
Here the function $x$~is the solution of~(\ref{dynamical_system_x_common_form})
with the same control~$c$ that is used in~(\ref{conjugate_system_ODE}). 
For the problem of minimizing $J_{M,2}(c)$:
\begin{align}
	p(T) = - 2 \left( \mathcal{F}(x(T); x_{\rm target}) - M\right) \beta \circ x_{\rm target}.
	\label{transversality_J_2_M_minimizing}
\end{align}
For the problem of minimizing $J^{\theta}_{M,1}(c)$:
\begin{align}
p(T) = \begin{cases}
	\beta \circ x_{\rm target}, & \mathcal{F}(x(T); x_{\rm target}) < M - \theta; \\
	-\beta \circ x_{\rm target}, & \mathcal{F}(x(T); x_{\rm target}) > M + \theta; \\	
	\frac{\mathcal{F}(x(T); x_{\rm target}) - M}{\theta} \beta \circ x_{\rm target}, 
	& \mathcal{F}(x(T); x_{\rm target}) \in [M - \theta, M + \theta]. \\	
\end{cases} 
\label{transversality_J_1_theta_M_minimizing}
\end{align} 

If only the constraints~(\ref{lower_constraints_to_incoherent_controls}) are used, then  
$c(t) \in Q_{\infty}:= \mathbb{R} \times [0, \infty)^2$.
Following the PMP theory, the Pontryagin function  should be maximized 
in the variables~$u, n_1, n_2$ for each~$t$. Thus, in addition to 
the constraint~(\ref{lower_constraints_to_incoherent_controls}), we consider the constraints 
$u_{\min}\le u(t) \le u_{\max}$, $n_j(t) \leq n_{\max}$, $\forall t \in [0, T]$, for some 
$u_{\min} < 0 < u_{\max}$, and $n_{\max} \in (0, \infty)$. Then $c(t) \in Q_{\rm compact}:= [n_{\min}, n_{\max}] \times [0, n_{\max}]^2$ for any $t \in [0, T]$. Correspondingly, we consider the functional classes $PC([0,T]; Q_{\infty})$ and $PC([0,T]; Q_{\rm compact})$ of piecewise continuous  controls. 

\begin{theorem} (PMP --- first-order necessary condition for optimality).\label{Theorem1}
Consider for the system~(\ref{dynamical_system_x_common_form}) with control 
$c = (u,~n_1,~n_2) \in PC([0,T]; Q_{\rm compact})$ 
and the problems of maximizing or minimizing~$J(c)$ and minimizing $J_{M,2}(c)$ or $J_{M,1}^{\theta}$, with fixed final time~$T$,  initial state~$x_0$ and target state $x_{\rm target}$. If~$c^{\ast}=(u^{\ast},n_1^{\ast},n_2^{\ast})$ is optimal control for some of these problems, then there exists a continuous vector function~$p^{\ast}$ which satisfies the conjugate system (\ref{conjugate_system_ODE}) with the corresponding  in~(\ref{transversality_J_min_or_max}--\ref{transversality_J_1_theta_M_minimizing}) transversality condition, where $u = u^{\ast}$, $n_1 = n_1^{\ast}$, $n_2 = n_2^{\ast}$, $x = x^{\ast}$, and the following maximization conditions are satisfied for any~$t \in [0, T]$:
\begin{align}
\max\limits_{u \in [u_{\min}, u_{\max}]} \mathcal{K}^u(p^{\ast}(t),x^{\ast}(t)) u  &= 
\mathcal{K}^u(p^{\ast}(t),x^{\ast}(t)) u^{\ast}(t), \label{maximization_condition_for_H_wrt_u} \\
\max\limits_{n_j \in [0, n_{\max}]} \mathcal{K}^{n_j} (p^{\ast}(t),x^{\ast}(t)) n_j &= 
\mathcal{K}^{n_j} (p^{\ast}(t),x^{\ast}(t)) n_j^{\ast}(t), \quad j = 1,2.
\label{maximization_condition_for_H_wrt_nj}
\end{align} 
Here $x^{\ast}$ is the solution of the system~(\ref{dynamical_system_x_common_form})
with control $c = c^{\ast}$.
\end{theorem} 

\subsection{Evolution of the Quantum System with Zero Controls}
\label{Subsection4.2}

If in (\ref{GKSL_eq}) coherent control $u=0$, then the operator $H_{u(t)} = V u(t) \equiv 0$ and the systems (\ref{GKSL_eq}), (\ref{dynamical_system_x_common_form})  
do not depend on the form of~$V$. For the control $c = \overline{c}$ with $\overline{u}= \overline{n}_1 = \overline{n}_2 = 0$,  
(\ref{dynamical_system_x_common_form}) becomes $\dot{\overline{x}} = A \overline{x}$, $\overline{x}(0)=x_0$
whose solution is $\overline{x} = e^{At}x_0$. 

Consider the following parameterized initial and target states:
\begin{align}
	& \rho_0 = {\rm diag}(a_1, a_2, a_3, a_4), \qquad
	\rho_{\rm target} = {\rm diag}(b_1, b_2, b_3, b_4)
	\label{parametrized_initial_and_target_density_matrices} \\ 
	& \text{s.t.} \qquad a_j, b_j \geq 0, \qquad j = 1,2,3,4, \qquad \sum\limits_{j=1}^4 a_j = 1, \qquad 
	\sum\limits_{j=1}^4 b_j = 1.
	\label{constraints_to_aj_bj}
\end{align}
The inequalities provide positive semi-definiteness and unit trace of the matrices. For~(\ref{parametrized_initial_and_target_density_matrices}), 
one has
\begin{align}
	x_0 &= (a_1,~\text{six zeros},~a_2,~\text{four zeros},~a_3,~0,~0,~a_4),
	\label{parametrized_initial_state_x_0} \\
	x_{\rm target} &=  (b_1,~\text{six zeros},~b_2,~\text{four zeros},~b_3,~0,~0,~b_4).
	\label{parametrized_target_state_x_target}
\end{align}
The solution of the system~(\ref{dynamical_system_x_common_form}) with (\ref{parametrized_initial_state_x_0}) and 
$c = \overline{c}$ is  
\begin{align}
\overline{x}_1 &=   a_1 + a_2 - a_2 e^{-2 \varepsilon\Omega_2 t} + e^{-2 \varepsilon(\Omega_1 + \Omega_2) t} 
(e^{ 2 \varepsilon\Omega_1 t} -1) (a_3 e^{2\varepsilon \Omega_2  t} + 
a_4 (e^{2 \varepsilon \Omega_2 t} - 1)), \nonumber \\
\overline{x}_8 &=   e^{-2  \varepsilon\Omega_2 t} (a_2 + a_4 - a_4 e^{-2 \varepsilon\Omega_1 t}), \quad
\overline{x}_{13} = e^{-2  \varepsilon \Omega_1 t} (a_3 + a_4 - a_4 e^{-2  \varepsilon \Omega_2 t}), \nonumber \\ 
\overline{x}_{16} &=   a_4 e^{-2 \varepsilon(\Omega_1 + \Omega_2) t}, \quad
\overline{x}_j = 0, \quad j \in \overline{1,16} \setminus \{1, 8, 13, 16\}. \qquad
\label{analytical_solution_quantum_system_x_under_zero_controls}
\end{align}

Consider the initial state $x_0 = $ (1, fifteen zeros), which corresponds to the pure state $\rho_0 = {\rm diag}(1,0,0,0)$. Then  the
system~(\ref{dynamical_system_x_common_form}) has the  solution
$\overline{x}_1(t) \equiv 1$ and $\overline{x}_i(t) \equiv 0$, $i=\overline{2,16}$.  The vector $A\overline{x}$ becomes zero, the system~(\ref{dynamical_system_x_common_form}) 
remains in this $x_0$ for an arbitrarily long time. The point $\overline{x} = (1,~\text{fifteen zeros})$ is singular for~(\ref{dynamical_system_x_common_form}). 

If~$\rho_0 \neq {\rm diag}(1,0,0,0)$, then for 
the system~(\ref{dynamical_system_x_common_form}) we have 
\begin{align}
	\lim\limits_{t\to\infty} \overline{\rho}(t) = 
	{\rm diag}\left(\lim\limits_{t\to\infty} \overline{x}_i(t)\right)_{i = 1, 8, 13, 16}
	= {\rm diag}\left(\sum\limits_{j=1}^4 a_j, 0, 0, 0 \right) 
	= {\rm diag}(1, 0, 0, 0). 
	\label{limit_gives_pure_state}
\end{align}  
The terminal function $\mathcal{F}(x; x_{\rm target})$ with $x = \overline{x}(t)$, $t \to \infty$ satisfies:
\begin{align}
	\lim\limits_{t \to \infty} \left\langle \overline{x}(t), \beta \circ x_{\rm target} \right\rangle = \lim\limits_{t \to \infty} \sum\limits_{i=1,8,13,16} b_i \overline{x}_i(t) = (a_1 + a_2 + a_3 + a_4) b_1.~~
	\label{limit_gives_certain_scalar_product}
\end{align}

\subsection{PMP. When Zero Coherent and Incoherent Controls Satisfy PMP}
\label{Subsection4.3}

For the system (\ref{dynamical_system_x_common_form}) with the initial
state (\ref{parametrized_initial_state_x_0}), consider
the problems of maximizing and minimizing $J(c)$ with $x_{\rm target}$ of the form~(\ref{parametrized_target_state_x_target}). Consider the conjugate system~(\ref{conjugate_system_ODE}, \ref{transversality_J_min_or_max}). For the control $c = \overline{c}$, this system becomes $\dot{\overline{p}} = -A^T \overline{p}$, $\overline{p}(T) = s \beta \circ x_{\rm target} = s x_{\rm target}$. Its solution is:
\begin{align*}
	\overline{p}_1 &=  b_1 s,\qquad 
	\overline{p}_8 = e^{-2 \varepsilon\Omega_2 T} \left( b_2 e^{2 \varepsilon\Omega_2 t} + 
	b_1 \left(-e^{2 \varepsilon\Omega_2 t} + e^{2 \varepsilon\Omega_2 T} \right) \right) s, \\
	\overline{p}_{13} &=  e^{-2 \varepsilon\Omega_1 T} \left( b_3 e^{2 \varepsilon\Omega_1 t} + 
	b_1 \left(-e^{2 \varepsilon\Omega_1 t} + e^{2 \varepsilon\Omega_1 T} \right) \right) s, \\
	\overline{p}_{16} &=  e^{-4 \varepsilon(\Omega_1 + \Omega_2) T} \big(b_2 \left(-e^{2 \varepsilon(\Omega_1 + \Omega_2) (t + T)} + e^{2 \varepsilon (2 \Omega_1 T + \Omega_2 t + \Omega_2 T)} \right) + e^{2 \varepsilon(\Omega_1 + \Omega_2) T} \big(-b_3 e^{2 \varepsilon(\Omega_1 + \Omega_2) t}\nonumber \\
	& \quad + b_4 e^{2 \varepsilon(\Omega_1 + \Omega_2) t} + b_3 e^{2 \varepsilon (\Omega_1 t  + \Omega_2 T)} + b_1 \big(e^{2 \varepsilon\Omega_1 t} - e^{2 \varepsilon\Omega_1 T}\big) 
	\big(e^{2 \varepsilon\Omega_2 t} - e^{2 \varepsilon\Omega_2 T} \big) \big)\big) s, \nonumber \\ 
	\overline{p}_j &=  0, \quad j \in \overline{1,16} \setminus \{1, 8, 13, 16\}.
\end{align*}
This system does not depend on $a_j$, $j=1,2,3,4$ 
and, because $u = 0$, on $V$.  

For any initial state $x_0$ of the form~(\ref{parametrized_initial_state_x_0}), 
$s \in \{\pm 1\}$, the system's parameters ($\Omega_1$, etc.), and~$T>0$, the switching function 
$\mathcal{K}^u$ is 
\begin{align}
	\mathcal{K}^u(\overline{x}(t), \overline{p}(t)) &=  0 \qquad \forall t \in [0, T]. \label{switching_function_wrt_u_is_zero} 
\end{align} 
In this sense, the control $u = \overline{u} = 0$ as a component of $\overline{c}$ is singular.  

Consider the two cases of $\rho_0$:
1)~$\rho_0 = {\rm diag}(1,0,0,0)$, i.e. $a_1 = 1$, $a_2 = a_3 = a_4 = 0$ in (\ref{parametrized_initial_state_x_0}) (for $u = n_1 = n_2 = 0$, the corresponding point $x_0 = (1,~\text{fifteen zeros})$ is singular for (\ref{dynamical_system_x_common_form})); 2)~$\rho_0 = \frac{1}{4}\mathbb{I}_4$ (completely mixed quantum state), i.e. $a_1 = a_2 = a_3 = a_4 = \frac{1}{4}$ in~(\ref{parametrized_initial_state_x_0}). 

A) Case when $\rho_0 = {\rm diag}(1,0,0,0)$. 
For zero controls, one gets that the  
functions
$\overline{x}_1 = 1$ and $\overline{x}_i = 0$, $i=\overline{2,16}$. The switching functions  become 
\begin{align}
	\mathcal{K}^{n_1}(\overline{x}(t), \overline{p}(t)) &= -2 (b_1 - b_3) s e^{2 \varepsilon\Omega_1 (t - T)} \varepsilon\Omega_1, 
	\label{switching_function_wrt_n1_case_rho0IsPure}  \\ 
	\mathcal{K}^{n_2}(\overline{x}(t), \overline{p}(t)) &= -2 (b_1 - b_2) s e^{2 \varepsilon\Omega_2 (t - T)} \varepsilon\Omega_2. 
	\label{switching_function_wrt_n2_case_rho0IsPure}
\end{align}
For (\ref{maximization_condition_for_H_wrt_u},\ref{maximization_condition_for_H_wrt_nj}), where instead of
$u^{\ast}(t)$, $n_i^{\ast}(t)$, $x^{\ast}(t)$, $p^{\ast}(t)$ we consider
$\overline{u}(t)$, $\overline{n}_i(t)$, $\overline{x}(t)$, $\overline{p}(t)$, 
our goal is to formulate sufficient conditions for
$b_1, b_2, b_3, b_4$ satisfying~(\ref{constraints_to_aj_bj}) to make the switching functions non-positive
for any $t\in[0, T]$ with any~$T \in (0, \infty)$. Thus, for (\ref{switching_function_wrt_n1_case_rho0IsPure}, \ref{switching_function_wrt_n2_case_rho0IsPure})
consider the algebraic system 
\begin{align}
	(b_1 - b_3) s \geq 0, \qquad (b_1 - b_2) s \geq 0 \qquad \text{s.t.} \quad (\ref{constraints_to_aj_bj}).
	\label{algebraic_system_for_case_rhoPure}
\end{align}
This system is solved via the function Reduce in Wolfram Mathematica.
For $s=1$ (the problem of maximizing the overlap), the logical condition is 
\begin{align}
	\big(\big(b_1 = 0 \wedge b_2 = 0 \wedge b_3 = 0\big) \lor \big(0 < b_1 \leq \frac{1}{3} \wedge 0 \leq b_2 \leq b_1 \wedge 0 \leq b_3 \leq b_1 \big) \nonumber \\
	\lor \big(\frac{1}{3} < b_1 < \frac{1}{2} \wedge b_3 \geq 0 \wedge \big(\big(2 b_1 + b_2 > 1 \wedge b_1 \geq b_2 \wedge b_1 + b_2 + b_3 \leq 1 \big) \nonumber \\
	\lor \big(b_1 \geq b_3 \wedge b_2 \geq 0 \wedge 2 b_1 + b_2 \leq 1\big)\big)\big) \lor \big(\frac{1}{2} \leq b_1 \leq 1 \nonumber \\
	\wedge \big(\big(b_2 \geq 0 \wedge b_1 + b_2 < 1 \wedge b_3 \geq 0 \wedge b_1 + b_2 + b_3 \leq 1 \big) \nonumber \\
	\lor \big(b_1 + b_2 = 1 \wedge b_3 = 0 \big)\big)\big)\big) \wedge b_1 + b_2 + b_3 + b_4 = 1.~~
	\label{compound_logical_condition_sIs1_rho0Pure}
\end{align}
Here ``$\wedge$'' is logical ``AND'' and ``$\lor$'' is logical ''OR''. 
These conditions provide non-positiveness of  
$\mathcal{K}^{n_i}$ for any $t\in[0, T]$ and, together 
with (\ref{switching_function_wrt_u_is_zero}), provide satisfaction of the PMP conditions   
(\ref{maximization_condition_for_H_wrt_u}, \ref{maximization_condition_for_H_wrt_nj}) in the class $PC([0,T]; Q_{\rm compact})$. This result does not depend on $V$ in $H_{u(t)} = V u(t)$.

For $s=-1$ (the problem of minimizing the overlap), we obtain from~(\ref{algebraic_system_for_case_rhoPure}): 
\begin{align}
\big(\big( b_1 = 0 \wedge \big(\big( 0 \leq b_2 < 1 \wedge b_3 \geq 0 \wedge b_2 + b_3 \leq 1 \big) \lor \big(b_2 = 1 \wedge b_3 = 0 \big)\big)\big) \nonumber \\ \lor \big(0 < b_1 \leq \frac{1}{3} \wedge \big(\big(b_1 \leq b_2 \wedge 2 b_1 + b_2 < 1 \wedge b_1 \leq b_3 \wedge b_1 + b_2 + b_3 \leq 1\big)  \nonumber \\
\lor \big(2 b_1 + b_2 = 1 \wedge b_1 + b_2 + b_3 = 1 \big)\big)\big)\big) \wedge b_1 + b_2 + b_3 + b_4 = 1.~~
\label{compound_logical_condition_sIsMinus1_rho0Pure}
\end{align}
These conditions in combination with (\ref{switching_function_wrt_u_is_zero}) provide satisfaction of   (\ref{maximization_condition_for_H_wrt_u}, \ref{maximization_condition_for_H_wrt_nj}).

B) Case when $\rho_0 = \frac{1}{4}\mathbb{I}_4$. The switching functions  $\mathcal{K}^{n_j}$, $j=1,2$ become 
\begin{align}
	\mathcal{K}^{n_1}(\overline{x}(t), \overline{p}(t)) &= - \left(e^{2 \varepsilon \Omega_1 t} - 1 \right)  
	\big(\big(2 e^{2 \varepsilon \Omega_2 T} - 1 \big)(b_1 - b_3) + b_2 - b_4 \big) s \Omega_1 E, \qquad
	\label{switching_function_wrt_n1_case_rho0IsMixed}  \\ 
	\mathcal{K}^{n_2}(\overline{x}(t), \overline{p}(t)) &= 
	-\left(e^{2 \varepsilon \Omega_2 t} - 1 \right)  
	\big(\big(2 e^{2 \varepsilon \Omega_1 T} - 1 \big)(b_1 - b_2) + b_3 - b_4 \big) s \Omega_2 E, \qquad 
	\label{switching_function_wrt_n2_case_rho0IsMixed}
\end{align}
where $E = \varepsilon e^{-2 \varepsilon (\Omega_1 + \Omega_2) T}$. Our goal is to make $\mathcal{K}^{n_j} \leq 0$, $j = 1,2$ for any $t\in[0,T]$. There is freedom of choice how to use~(\ref{switching_function_wrt_n1_case_rho0IsMixed},\ref{switching_function_wrt_n2_case_rho0IsMixed}) for constructing the corresponding inequality constraints. For $s = 1$, if we consider the system
\begin{align*}
	b_1 - b_3 \geq 0, \quad b_2 - b_4 \geq 0,
	\quad b_1 - b_2 \geq 0, \quad b_3 - b_4 \geq 0
	\qquad \text{s.t.} \quad (\ref{constraints_to_aj_bj}),
\end{align*}
then the explicit solution is derived, but the result is quite cumbersome. Therefore, for $s=1$ let 
take the following simpler algebraic system:
\begin{align}
	b_1 - b_3 = 0, \quad b_2 - b_4 \geq 0,
	\quad b_1 - b_2 = 0, \quad b_3 - b_4 \geq 0
	\qquad \text{s.t.} \quad (\ref{constraints_to_aj_bj})
	\label{algebraic_system_for_case_rhoMixed_s1}
\end{align}
whose solution is
\begin{align}
\frac{1}{4} \leq b_1 \leq \frac{1}{3} \wedge b_1 = b_2 \wedge b_1 = b_3 \wedge b_1 + b_2 + b_3 + b_4 = 1.
\label{compound_logical_condition_rho0Mixed_s1}
\end{align}
Here the problem of maximizing $J(c)$ describes partial purification of the system's final state. 
By analogy with (\ref{algebraic_system_for_case_rhoMixed_s1}), for $s=-1$ consider the system
\begin{align*}
	b_1 - b_3 = 0, \quad b_2 - b_4 \leq 0,
	\quad b_1 - b_2 = 0, \quad b_3 - b_4 \leq 0
	\qquad \text{s.t.} \quad (\ref{constraints_to_aj_bj})
\end{align*}
whose solution is 
\begin{align}
	0 \leq b_1 \leq \frac{1}{4} \wedge b_1 = b_2 \wedge b_1 = b_3 \wedge b_1 + b_2 + b_3 + b_4 = 1.
	\label{compound_logical_condition_rho0Mixed_sMinus1}
\end{align}

C) The obtained above results about the conditions under which the control $c = \overline{c} = 0$ satisfies the PMP (Theorem~\ref{Theorem1}), for the two cases of~$\rho_0$ are  summarised as the next theorem formulated in the terms of density matrices. 
\begin{theorem} 
\label{Theorem2} Consider the system (\ref{GKSL_eq}) with control $c \in PC([0,T]; Q_{\rm compact})$ and with the two cases of initial state $\rho_0$: (1)~pure $\rho_0 = {\rm diag}(1,0,0,0)$; (2) completely mixed $\rho_0 = \frac{1}{4}\mathbb{I}_4$. Consider the problems of maximizing and minimizing~$J(c)$ with the target state $\rho_{\rm target} = {\rm diag}(b)$, where $b = (b_1, b_2, b_3, b_4)$. Then the control $c = \overline{c} = 0$ satisfies the PMP in the class $PC([0,T]; Q_{\rm compact})$ for any arbitrary given Hermitian~$V$, system's parameters ($\Omega_1 > 0$, etc.), $u_{\min} < 0 < u_{\max}$ and $n_{\max} > 0$, any $T>0$, if vector $b$ is as follows:
\begin{itemize}
    \item for $\rho_0 = {\rm diag}(1,0,0,0)$: for the problem of maximizing $J(c)$ if $b$ satisfies~(\ref{compound_logical_condition_sIs1_rho0Pure});
    for the problem of minimizing $J(c)$ if $b$ satisfies~(\ref{compound_logical_condition_sIsMinus1_rho0Pure});
    \item for $\rho_0 = \frac{1}{4}\mathbb{I}_4$: for the problem of maximizing $J(c)$ if $b$ satisfies~(\ref{compound_logical_condition_rho0Mixed_s1});
    for the problem of minimizing $J(c)$ if $b$ satisfies~(\ref{compound_logical_condition_rho0Mixed_sMinus1}).
\end{itemize}
\end{theorem}

\subsection{Lower and Upper Bounds for the Overlap} 
\label{subsection4.4}

In the general theory of optimal control, an important problem is to establish 
whether a~control that satisfies the~PMP is indeed globally optimal. For the objective $\langle \rho(T), \rho_{\rm target} \rangle$ the lower and upper bounds are obviously known as the extremal values in the finite-dimensional optimization problems
$\langle \rho, \rho_{\rm target} \rangle \to \min$
and $\langle \rho, \rho_{\rm target} \rangle \to \max$, correspondingly, over all density matrices, as e.g. in~\cite[p.~2]{Oza_Pechen_Dominy_Beltrani_Moore_Rabitz_2009}) and are discussed below in our terms.
 
For a diagonal target density matrix 
$\rho_{\rm target} = {\rm diag}(b_1, b_2, b_3, b_4)$, one has  
$\langle \rho, \rho_{\rm target} \rangle = \sum\limits_{j=1}^4 b_j \rho_{j,j}$ and $
\mathcal{F}(x; x_{\rm target}) =\langle x, \beta \circ x_{\rm target} \rangle = \sum\limits_{j=1,8,13,16} x_j x_{{\rm target},j} 
\to \min$ or $\max$ s.t.~(\ref{trace_condition}) and $x_j \geq 0$, $j = 1,~8,~13,~16$. Hence for the diagonal case the minimal and maximal values
of the overlap $\langle \rho, \rho_{\rm target} \rangle$ are equal to the minimal and maximal eigenvalues of $\rho_{\rm target}$, correspondingly. The problem of maximizing the overlap $\langle \rho, \rho_{\rm target} \rangle$
with a non-diagonal $\rho_{\rm target}$ is reduced
to the diagonal case via  diagonalization
$\rho_{\rm target} = U \rho_{\rm target}^{\rm diag} U^{\dagger}$ with 
diagonal $\rho_{\rm target}^{\rm diag}$ and unitary $U$, and using the equalities 
$\max\limits_{\rho} {\rm Tr}(\rho \rho_{\rm target}) = 
\max\limits_{\rho} {\rm Tr}(\rho U \rho_{\rm target}^{\rm diag} U^{\dagger}) =  
\max\limits_{\rho} {\rm Tr}(U^{\dagger} \rho U \rho_{\rm target}^{\rm diag}) 
= \max\limits_{\rho} {\rm Tr}(\rho \rho_{\rm target}^{\rm diag})$  
where the last equality holds because the set of all density matrices
is invariant under the unitary transformation $\rho \to  U^{\dagger} \rho U$. Similar analysis can be done for the problem of minimizing the overlap. Thus, for general $\rho_{\rm target}$ the minimal and maximal bounds 
of the objective can be chosen as the minimal and maximal eigenvalues of $\rho_{\rm target}$.

One can expect a~case when for a given final time~$T$ it is not possible to at least approximately reach such an bound (lower or upper) with any control from the class $PC([0,T]; Q_{\infty})$ or $PC([0,T]; Q_{\rm compact})$. However, if one maximizes the objective with some $\rho_{\rm target}$ and $T$ and obtains that the objective reaches the upper bound with a good precision, then it allows to conclude that the maximization  problem is globally solved at least approximately. 

\subsection{When Zero Coherent and Incoherent Controls Form a~Stationary Point}
\label{Subsection4.5}

An admissible control $\widetilde{c} = (\widetilde{u}, \widetilde{n}_1, \widetilde{n}_2)$ is called {\it first-order singular} (we also write ``singular'') \cite{GabasovBook1978} at some subset $\Omega \subseteq [0, T]$ with ${\rm mes}\,\Omega > 0$, if  $h(\widetilde{x}(t), \widetilde{p}(t), u, n_1, n_2)$ does not depend on the variables $u$, $n_1$, $n_2$ at $\Omega$, i.e. $h(\widetilde{x}(t), \widetilde{p}(t), u, n_1, n_2) - h(\widetilde{x}(t), \widetilde{p}(t), \widetilde{u}(t), 
\widetilde{n}_1(t), \widetilde{n}_2(t)) \equiv 0$, $t \in \Omega$. Because the function $h$ is linear in $u,~n_j$, this definition means that 
$\mathcal{K}^u (u(t) - \widetilde{u}(t)) + \sum\limits_{j=1}^2 \mathcal{K}^{n_j} (n_j(t) - \widetilde{n}_j(t)) \equiv 0$
and all the functions $\mathcal{K}^u (\widetilde{x}(t), \widetilde{p}(t)) \equiv 0$, 
$\mathcal{K}^{n_j}(\widetilde{x}(t), \widetilde{p}(t)) \equiv 0$. 
In other words, the function~$h(\widetilde{x}(t), \widetilde{p}(t), u, n_1, n_2)$ to be maximized with respect to the variables $u$, $n_1$, $n_2$ does not depend on these variables. Because the problem is  linear in controls, the definition of a~control being singular at the whole time range (i.e. for $\Omega = [0,T]$) coincides with the definition of
a stationary point of $J(c)$ in terms of the switching functions.

Set $\rho_0 = {\rm diag}(1,0,0,0)$. Below we find the conditions which provide all the switching functions to be zero at the whole~$[0, T]$ for the control~$c = \overline{c}$. Using (\ref{algebraic_system_for_case_rhoPure}), consider the algebraic system $b_1 - b_3 = 0$, $b_1 - b_2 = 0$ s.t.  (\ref{constraints_to_aj_bj}). Solving this system, we derive the following compound condition for both $s = \pm 1$:
\begin{align}
	0 \leq b_1 \leq \frac{1}{3}~\wedge~b_1 = b_2 = b_3~\wedge~b_4 = 1 - 3 b_1.
	\label{Conditions_when_all_zero_controls_u_n_are_singular_rho0Pure}
\end{align}
Then using~(\ref{switching_function_wrt_u_is_zero}) we obtain the next theorem in terms of density matrices.

\begin{theorem} (Zero control as a stationary point of $J(c)$). \label{Theorem3} Consider the system (\ref{GKSL_eq}) with $\rho_0 = {\rm diag}(1,0,0,0)$, $c \in PC([0,T]; Q_{\rm compact})$ or $c \in PC([0,T]; Q_{\infty})$. 
Consider the problems of maximizing and minimizing~$J(c)$ with the target state $\rho_{\rm target} = {\rm diag}(b_1, b_2, b_3, b_4)$. Then the control $c = \overline{c} = 0$ is a stationary point of $J(c)$
in $PC([0,T]; Q_{\rm compact}) \subset PC([0,T]; Q_{\infty})$ for any given Hermitian~$V$, system's parameters ($\Omega_1 > 0$, etc.), $u_{\min} < 0 < u_{\max}$ and $n_{\max} > 0$ (these bounds are used if $Q = Q_{\rm compact}$), any $T>0$, if vector $(b_1, b_2, b_3, b_4)$ satisfies the condition~(\ref{Conditions_when_all_zero_controls_u_n_are_singular_rho0Pure}). 
\end{theorem}

Comparing Theorems~\ref{Theorem2} and~\ref{Theorem3} allows to show the effect from the controls' constraints. One can expect  a~case when the control $c = \overline{c} = 0$ is not a~stationary point of $J(c)$, but, following Theorem~\ref{Theorem2}, it may turn out that this control is a~candidate for being optimal.

\subsection{Analytical Study When Zero Singular Coherent and Incoherent Controls Are 
	Exactly Optimal} 
\label{Subsection4.6}

In regard to Theorems~\ref{Theorem2},~\ref{Theorem3}, consider $\rho_0 = {\rm diag}(1,0,0,0)$ and $\rho_{\rm target} = \frac{1}{5}{\rm diag}(1,1,1,2)$ which satisfy (\ref{Conditions_when_all_zero_controls_u_n_are_singular_rho0Pure}), i.e. the control $\overline{c} = 0$ is a stationary point of $J(c)$. As (\ref{analytical_solution_quantum_system_x_under_zero_controls}) shows, $\overline{c}$ gives  $\overline{x}_1(t) \equiv a_1 = 1$, $\overline{x}_j(t) \equiv 0$, $j = \overline{2,16}$ and, therefore, we analytically obtain $J(\overline{c}) = \langle x(T), \beta \circ x_{\rm target} \rangle = a_1 x_{{\rm target},1} = b_1 = 1/5$ for any~$T$. For this $\rho_{\rm target}$, we using Subsect.~\ref{subsection4.4} obtain that the 
lower and upper bounds of $J(c)$ are 1/5 and 2/5, correspondingly. 

For the problem of minimizing $J(c)$ we obtain that  the stationary point $\overline{c} = 0$ provides $J(\overline{c}) = 1/5$ that is exactly equal to the upper bound for any $T > 0$,  Hermitian~$V$,  etc. This means that $\overline{c} = 0$ is exactly globally optimal, $J(\overline{c})$ is the globally minimal value of~$J(c)$.

For the problem of maximizing $J(c)$ we see that the value $J(\overline{c})$ at the stationary point $\overline{c} = 0$ is two times smaller than the upper bound for any~$T$. We perform the following numerical illustration for $T = 2$. 
Set the controls $u = 10 \sin t$, $n_1 = n_2 = 0$, $t \in [0, 2]$, the values  
$\varepsilon = 0.1$, $\omega_1 = 1$, $\omega_2 = \Omega_j = 0.5$, $\Lambda_j = 0.05$, $j=1,2$. Solving numerically~(\ref{dynamical_system_x_common_form}), we obtain $J(c) \approx 0.37 > J(\overline{c}) = 0.2$. In our problem, PMP is not a sufficient condition for optimality.

\section{Gradient Projection Methods}
\label{Section5}

Consider the objective functional $I(c)$ to be minimized, where, as it is introduced in Sect.~\ref{Section3}, 
$I(c)$ is either $\overline{J}-J(c)$, 
or $J(c)$, or $J_{M,2}(c)$, or $J_{M,1}^{\theta}(c)$. In terms of sequential 
updates of controls, one sets an initial admissible guess 
$c = c^{(0)} = (u^{0)}, n_1^{(0)}, n_2^{(0)})$ and aims to construct such 
a~sequence $c^{(0)},~c^{(1)},~\dots,~c^{(k)},~\dots,~c^{(M)}$ that 
the corresponding values of $I(c)$ should either monotonically decrease ($I(c^{(k+1)}) < I(c^{(k)})$ 
or decrease in general. Consider a current control $c^{(k)}$, $k \geq 0$ 
and an arbitrary admissible  control~$c$. At them, 
using the corresponding
general result from the theory of optimal control~\cite{DemyanovBook1970, LevitinUSSRComputMathMathPhys1966}, 
we have the expansion
\begin{align}
I(c) - I(c^{(k)}) &=  \left\langle {\rm grad}\, I(c^{(k)}), c - c^{(k)} \right\rangle_{L^2} + r =  -\int\limits_0^T \Big[ \mathcal{K}^{u}(p^{(k)}(t), x^{(k)}(t)) (u(t)  - u^{(k)}(t)) \nonumber \\
& \quad + \sum\limits_{j=1}^2  \mathcal{K}^{n_j}(p^{(k)}(t), x^{(k)}(t)) (n_j(t) - n_j^{(k)}(t)) \Big] dt + r, \label{I_increment_formula} \\
{\rm grad}\, I(c^{(k)})(t) &=  - h_c(p,x,c)\big\vert_{p=p^{(k)}(t),~x=x^{(k)}(t),~c=c^{(k)}(t)} \nonumber \\
&= \left(\mathcal{K}^{u}(p^{(k)}(t), x^{(k)}(t)),~~ \mathcal{K}^{n_j}(p^{(k)}(t), x^{(k)}(t)), ~~ j = 1, 2 \right). \nonumber 
\end{align} 
where $r$ is the remainder, ${\rm grad}\, I(c^{(k)})$ is the gradient of $I(c)$ at $c^{(k)}$, and this gradient is the vector-function defined at the whole $[0,T]$.

For the problem of minimizing~$I(c)$,  
consider the following iterative process of 
GPM-2 starting from an admissible 
$c^{(0)}$: 
\begin{align}
c^{(k+1)}(t) = {\rm Pr}_Q\Big(c^{(k)}(t) - \alpha^{(k)}\, {\rm grad}\, I(c^{(k)})(t) + 
\beta \left(c^{(k)}(t) - c^{(k-1)}(t) \right) \Big), 
\label{GPM2_general_formula}
\end{align}
where $Q = Q_{\infty}$ or $Q = Q_{\rm compact}$; if $k=0$, then we use $\alpha^{(0)} > 0$, $\beta = 0$, and if $k>0$, then we use  $\alpha^{(k)} > 0$, $\beta \in (0, 1)$; the orthogonal projection ${\rm Pr}_Q$ maps any point outside of $Q$ to a closest
point in $Q$, and leaves unchanged points in~$Q$. 
In details, we have
\begin{align}
	u^{(1)}(t) &=  {\rm Pr}_{Q_u} \left(u^{(0)}(t) + \alpha^{(0)} \mathcal{K}^{u}(p^{(0)}(t), x^{(0)}(t)) \right),
	\label{GPM_f1} \\
	n_j^{(1)}(t) &= {\rm Pr}_{Q_{n_j}} \left(n_j^{(0)}(t) + \alpha^{(0)} \mathcal{K}^{n_j}(p^{(0)}(t), x^{(0)}(t)) \right),
	\label{GPM_f2} \\
	u^{(k+1)}(t) &=  {\rm Pr}_{Q_u}\Big(u^{(k)}(t) + \alpha^{(k)} \mathcal{K}^{u}(p^{(k)}(t), x^{(k)}(t))+ \beta(u^{(k)}(t) - u^{(k-1)}(t)) \Big), 
	\label{GPM_f3} \\
	n_j^{(k+1)}(t) &=  {\rm Pr}_{Q_{n_j}}\Big(n_j^{(k)}(t) + 
	\alpha^{(k)} \mathcal{K}^{n_j}(p^{(k)}(t), x^{(k)}(t))
	+ \beta(n_j^{(k)}(t) - n_j^{(k-1)}(t)) \Big), 
	\label{GPM_f4}
\end{align}
where $k \geq 1$; $j = 1,2$; $Q_u = \mathbb{R}$ or $Q_u = [u_{\min}, u_{\max}]$, $Q_{n_j} = [0, \infty)$ or $Q_{n_j} = [0, n_{\max}]$; the parameters $\alpha^{(k)} > 0$ ($k = 0, 1, 2, \dots$) and $\beta \in (0, 1)$ should be adjusted. E.g., one can try to fix $\alpha^{(k)} > 0$ to be appropriate for the whole set of iterations. An another rule is $\alpha^{(k)} = \widehat{\alpha}/(k^{\sigma}+1)$, where  $\widehat{\alpha},~\sigma > 0$. As~(\ref{GPM_f1}, \ref{GPM_f2}) show, 
the first iteration of GPM-2 coincides with the first iteration of GPM-1. Note the inertial terms $\beta(u^{(k)}(t) - u^{(k-1)}(t))$ and $\beta(n_j^{(k)}(t) - n_j^{(k-1)}(t))$ in (\ref{GPM_f3}, \ref{GPM_f4}). 
Also note that (\ref{GPM2_general_formula}) has the same sign ``$-$'' before the gradient both for the problems of minimizing and  maximizing, because the difference between these two problems is taken into account in~(\ref{transversality_J_min_or_max}). In general, we do not expect that GPM-1 and GPM-2 give $I(c^{(k+1)}) < I(c^{(k)})$ at each iteration. 

Consider the stopping criterion $\vert I(c^{(k+1)}) - I(c^{(k)}) \rvert < \varepsilon_{{\rm stop},1} \ll 1$. 
Together with this, for the problem of minimizing $I = J^{\theta}_{M,1}(c)$, additionally consider the stopping criterion $J^{\theta}_{M,1} < \varepsilon_{{\rm stop},2} \ll 1$. For minimizing $I = J^{\theta}_{M,1}(c)$, we use $J_{M,1}(c)$ in the stopping criterion $J_{M,1}(c) < \varepsilon_{{\rm stop},3}$.  

By the construction, 
$\frac{1}{T}\int\limits_0^T \sum\limits_{j=1}^4 \rho_{j,j}(t)dt = 
\frac{1}{T}\int\limits_0^T \sum\limits_{j =1,8,13,16} x_j(t)dt = 1$. Set 
\begin{align*}
	\aleph & := \frac{1}{T} \int\limits_0^T   
 \sum\limits_{i,j =1,2,3,4,~ i<j}
 \lvert\rho_{i,j}(t)\rvert^2  dt = \frac{1}{T} \int\limits_0^T 
	\sum\limits_{j = \overline{2,15}\setminus{\{8,13\}}} x_j^2(t) dt  \nonumber \\
&  \approx \frac{1}{T} \sum\limits_{j \in \overline{2,15}\setminus{\{8,13\}}} x_j^2(t_q) \Delta t 
	= \frac{1}{K} \sum\limits_{j\in \overline{2,15}\setminus{\{8,13\}}} x_j^2(t_q),
\end{align*}
where $\Delta t = T/K$ with a sufficiently large $K \in \mathbb{N}$.

\section{Numerical Results with GPM-1 and GPM-2}
\label{Section6}

Consider either $V = V_1$ or $V= V_2$ defined in 
(\ref{V_variants_1_and_2}). As in Subsect.~\ref{Subsection4.6}, use $\varepsilon = 0.1$, $\omega_1 = 1$, $\omega_2 = \Omega_j = 0.5$, $\Lambda_j = 0.05$, $j=1,2$. The described below numerical experiments were carried out without reduction to finite-dimensional optimization. They use our implementation of GPM in Python using  {\tt solve\_ivp} from {\tt SciPy}, etc. As an option, take $Q = Q_{\infty}$. In the stopping criteria, take $\varepsilon_{{\rm stop},1} = 10^{-8}$ (in Subsect.~\ref{Subsection6.1}--\ref{Subsection6.3}) and 
$\varepsilon_{{\rm stop},j} = 10^{-4}$, $j=2,3$ (in Subsect.~\ref{Subsection6.3}). It is interesting to analyze various mathematical aspects, e.g.,   
convergence of $\{ c^{(k)} \}$ to zero in Subsect.~\ref{Subsection6.1},~\ref{Subsection6.2}
with respect to Theorem~\ref{Theorem2}. 

\subsection{Maximizing the Overlap for Mixed $\rho_{\rm target}$} 
\label{Subsection6.1}

\begin{figure}[ht!]
\centering
\includegraphics[width=1\linewidth]{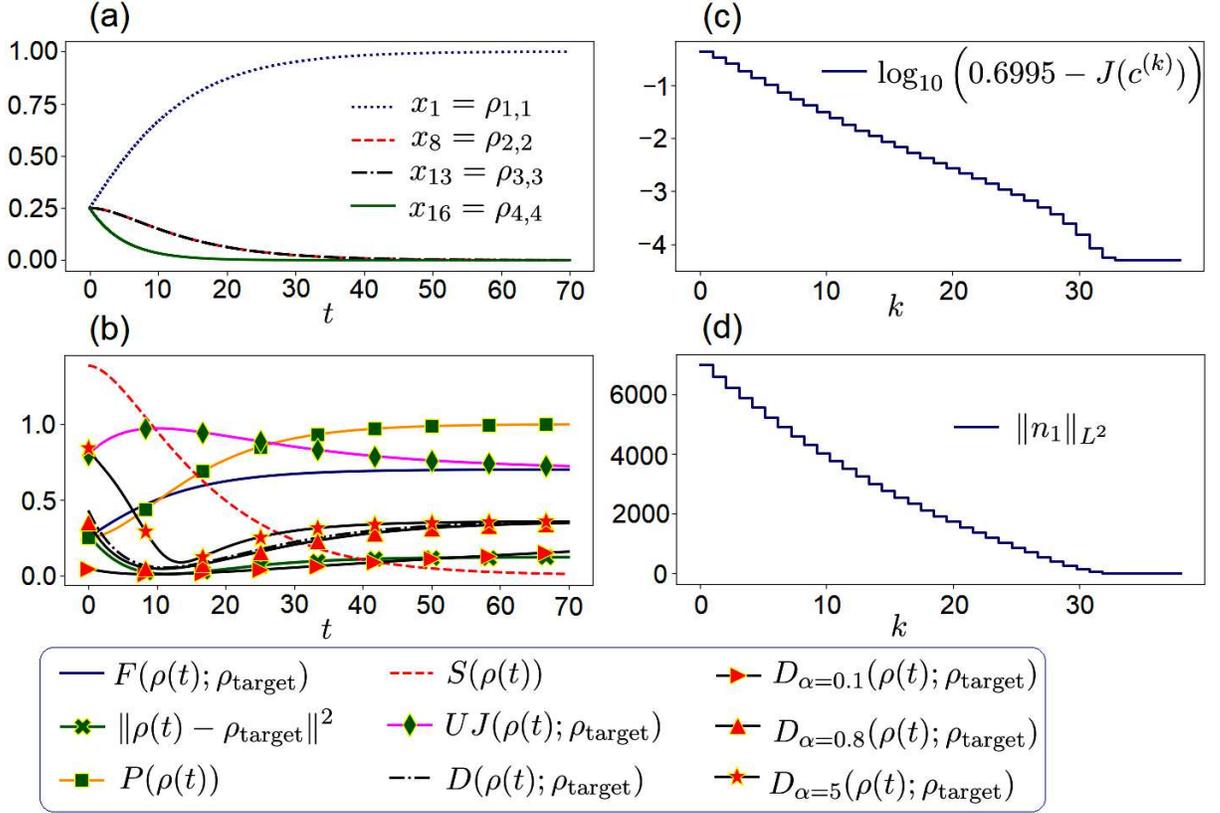} 
\caption{The results of GPM-2 for the problem of maximizing $J(c)$ with $\rho_0 = \frac{1}{4}\mathbb{I}_4$ and $\rho_{\rm target} = \frac{1}{10}{\rm diag}(7, 1, 1, 1)$:
    (a)~the functions $\rho_{j,j}$ ($j =1,2,3,4$) of~$t$, where $\rho_{1,1}\approx 1$ and   $\rho_{j,j} \approx 0$ ($j=2,3,4$) at $t=T$;
    (b)~evolution at the optimized $c$ of: the overlap $F(\rho(t); \rho_{\rm target})$ reaching the upper bound 0.7 with a~good precision at~$t=T$, squared distance $\| \rho(t) - \rho_{\rm target} \|^2$, purity reaching $\approx 0.998$ at~$t=T$, von Neumann entropy reaching $\approx 0.008$ at~$t=T$, Uhlmann--Jozsa fidelity, quantum relative entropy, Petz--R\'{e}nyi relative entropy with $\alpha = 0.1,~0.8,~5$;
    (c)~monotonous decrease of~$I$ versus the index~$k$ over the iterations of GPM-2 
    in the $\log_{10}$ scale;
    (d)~monotonous decrease of $L^2$-norm for $n_1$ versus~$k$.}
\label{Fig2}
\end{figure}

Consider $\rho_0 = \frac{1}{4}\mathbb{I}_4$ and $\rho_{\rm target} = \frac{1}{10}{\rm diag}(7, 1, 1, 1)$. The entropy $S(\rho_{\rm target}) \approx 0.94$ is about 68~\% of the entropy of the completely mixed state~$\rho_0$. 
Consider $V = V_1$. For this $\rho_{\rm target}$, using Subsect.~\ref{subsection4.4} we obtain that the upper bound for $J(c)$ is 0.7. For this $\rho_0 = \frac{1}{4}\mathbb{I}_4$, $c = \overline{c} = 0$, and $T = 70$, using (\ref{analytical_solution_quantum_system_x_under_zero_controls}) we compute that $J(\overline{c}) \approx 0.6994$, i.e. sufficiently close to~0.7. This is consistent with~(\ref{limit_gives_pure_state}) 
and~(\ref{limit_gives_certain_scalar_product}). For the problem of maximizing $J(c)$, we use the objective $I(c) := 0.6995 - J(c)$ to be minimized.  

Take $T = 70$. Set $c^{(0)}$ formed by $n_1^{(0)} = n_1^{(0)} = 10$, $u^{(0)} = 0$ that give $I(c^{(0)}) \approx 0.43$. We use GPM-1, GPM-2 with this $c^{(0)}$ and fixing $\alpha^{(k)} = 10^5$, $\beta = 0.9$. For computing needs, consider the uniform time grid with the step $T/N$ with $N=1000$.  Piecewise constant interpolation 
for $u,~n_j$ is used. After solving 73 Cauchy problems (for (\ref{dynamical_system_x_common_form},   \ref{conjugate_system_ODE})), GPM-2 provides $I \approx 6 \cdot 10^{-4}$, i.e. near 0.1~\% of~$I(c^{(0)})$, $\aleph \approx 0$. Fig.~\ref{Fig2} shows the results obtained with GPM-2. As Fig.~\ref{Fig2}(a) shows, $\rho(T)$ is not close to $\rho_{\rm target}$. For comparing, use GPM-1 (i.e. $\beta = 0$ for all~$k$ in (\ref{GPM2_general_formula})) with $\alpha^{(k)} = 10^5$ for all $k$. For the same $\varepsilon_{{\rm stop},1}$, GPM-1 gives also $I \approx 6 \cdot 10^{-4}$ at the cost of 87 Cauchy problems. Thus, GPM-2 with $\beta = 0.9$ is faster than GPM-1 for the same fixed $\alpha^{(k)}$. 

\subsection{Maximizing the Overlap for Pure \texorpdfstring{$\rho_{\rm target}$}{the Target Density Matrix}} 
\label{Subsection6.2}

Consider $\rho_0 = \frac{1}{4}\mathbb{I}_4$, $V = V_1$. Instead of $\rho_{\rm target} = {\rm diag}(0.7, 0.1, 0.1, 0.1)$ whose von Neumann entropy is 0.94, consider $\rho_{\rm target} = {\rm diag}(1,0,0,0)$ whose von Neumann entropy is zero (as the linear entropy $1-P(\rho_{\rm target})$). For this $\rho_{\rm target}$, using Subsect.~\ref{subsection4.4} we compute that the 
upper bound for $J(c)$ is~1 (instead of 0.7). This is consistent  with~(\ref{limit_gives_pure_state}) and~(\ref{limit_gives_certain_scalar_product}).

\begin{figure}[ht!]
	\centering
	\includegraphics[width=1\linewidth]{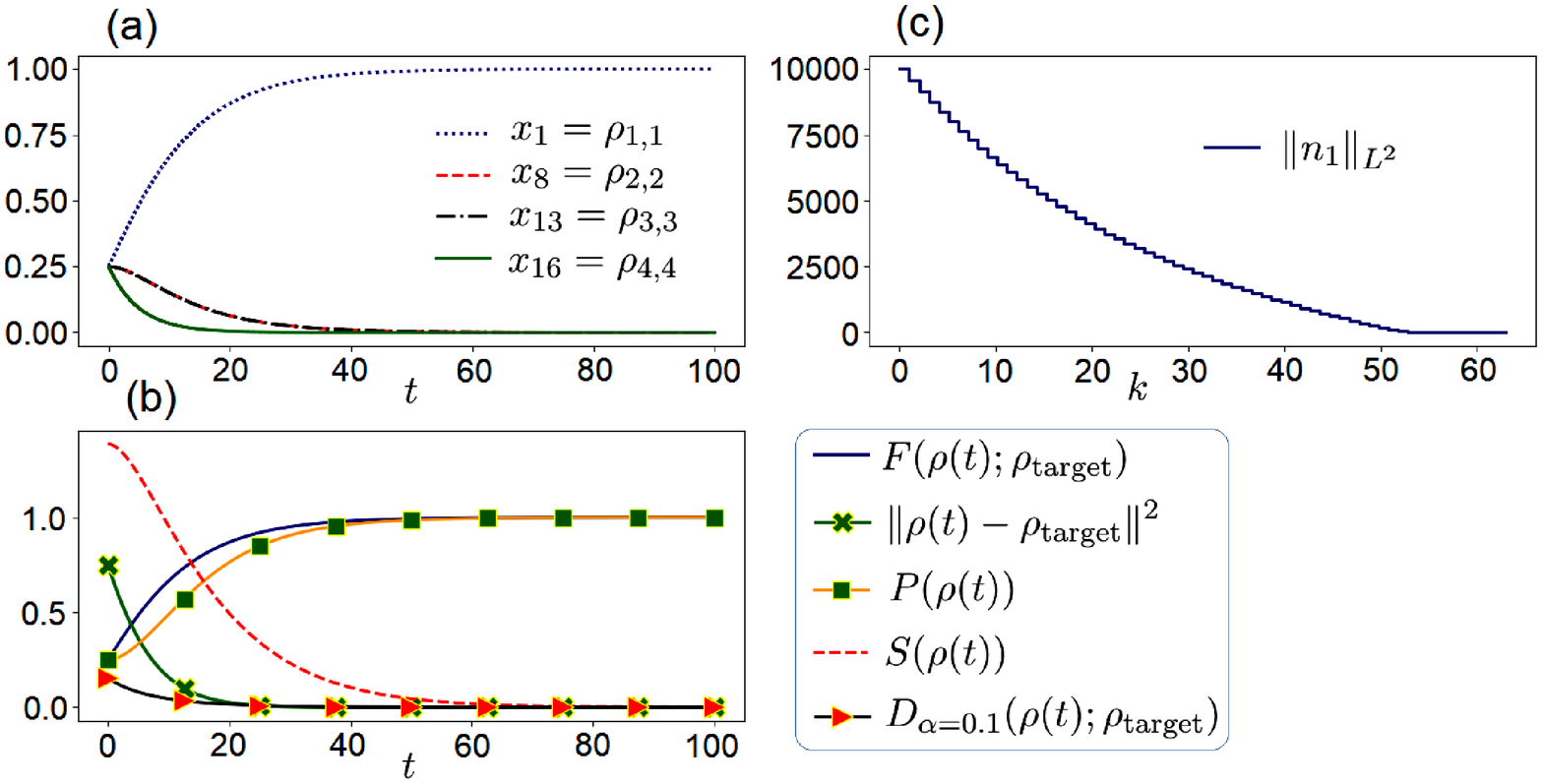} 
	\caption{The results of the GPM-2
		for maximizing the overlap $\langle\rho(T), \rho_{\rm target} \rangle$
		with maximally mixed $\rho_0 = \frac{1}{4}\mathbb{I}_4$ and pure state
		$\rho_{\rm target} = {\rm diag}(1,0,0,0)$. }
	\label{Fig3}
\end{figure}

\begin{table}[ht!]
\begin{center}
\begin{minipage}{\textwidth}
\caption{About the work of GPM-2 for the problems of maximizing the overlap for the two variants of $(\rho_{\rm target}, T)$ (Subsect.~\ref{Subsection6.1},~\ref{Subsection6.2})}\label{Table1}
\begin{tabular*}{\textwidth}{@{\extracolsep{\fill}}lcc@{\extracolsep{\fill}}}
\hline 
 \backslashbox[0pt][l]{Data and Results}{Case $(\rho_{\rm target},T)$} & $(\text{mixed}, 70)$ & $(\text{pure}, 100)$ \\\hline
    $(J(c^{(0)}), I(c^{(0)}))$ \--- initial values & $\approx (0.27, 0.43)$ & $\approx (0.27, 0.73)$ \vspace{0.2cm}\\  
    $\overline{J}$ for defining $I(c) = \overline{J} - J(c)$ & 0.6995 & 1 \vspace{0.2cm}\\   
    $(\alpha^{(k)}, \beta)$ & $(10^5, 0.9)$ &  $(10^5, 0.9)$ \vspace{0.2cm} \\ 
    \begin{tabular}{@{}l@{}}Number of solving the Cauchy  problems \end{tabular} & 73 & 119 \vspace{0.2cm}\\   
    $I(c)$ at the obtained $c$ & $\approx 6 \cdot 10^{-4}$ &  $\approx 5 \cdot 10^{-5}$ \vspace{0.2cm} \\  
    $\aleph$ at the obtained $c$ & $\approx 0$ & $\approx 0$ \vspace{0.2cm} \\  
    Corresponding Fig. and Subsec. &  Fig.~\ref{Fig2},  Subsect.~\ref{Subsection6.1}   
         &  Fig.~\ref{Fig3},  Subsect.~\ref{Subsection6.2}  \vspace{0.2cm} \\  
   $S(\rho(T)) \approx 0$ at the obtained $c$ & Yes & Yes \vspace{0.2cm} \\  
   $\langle \rho(T), \rho_{\rm target} \rangle \approx 1$ at the obtained $c$ & No & Yes \vspace{0.2cm} \\  
\hline
\end{tabular*}
\end{minipage}
\end{center}
\end{table}

Define $I(c) = 1 - J(c)$, take $T = 100$. For storing controls in computer, $[0, T]$ is divided into $N=1000$ pieces of the equal length. Set 
$n_1^{(0)} = n_1^{(0)} = 10$ and $u^{(0)} = 0$  
that give $J(c^{(0)}) \approx 0.27$, $I(c^{(0)}) \approx 0.73$.
We use GPM-1, GPM-2 with this $c^{(0)}$ and fixing  $\alpha^{(k)} = 10^5$, $\beta = 0.9$. At the cost of solving 119 Cauchy problems, GPM-2 gives $I(c) \approx 5 \cdot 10^{-5}$, $\aleph \approx 0$. Fig.~\ref{Fig3}(a) shows the numerically optimized functions $\rho_{j,j}$ ($j=1,2,3,4$) which are similar to the functions shown in~Fig.~\ref{Fig2}(a). In contrast to Fig.~\ref{Fig2}(b), we see in~Fig.~\ref{Fig3}(b) that $F(\rho(t); \rho_{\rm target}) \approx 1$ at $t = T$ and 
$\| \rho(t) - \rho_{\rm target}\|^2 \approx 0$ at $t = T$. We mean that, in contrast to the previous case with non-pure $\rho_{\rm target}$, here the overlap reaches approximately~1 and the distance reaches approximately~0 at the appropriate~$T$. Here maximizing $\langle \rho(T), \rho_{\rm target} \rangle$ gives also approximate steering $\rho_0 \to \rho_{\rm target}$. For comparing, we use GPM-1 with the same $\alpha^{(k)}$. At the cost of solving 141  Cauchy problems, GPM-1 gives $I(c) \approx 5 \cdot 10^{-5}$, i.e. GPM-2 with $\beta = 0.9$ is faster than GPM-1. 

Table~\ref{Table1} summarizes the information about the work of GPM-2 from  Subsect.~\ref{Subsection6.1},~\ref{Subsection6.2}. With respect to the noticeably different resulting values of $I$ in these cases, note that the stopping criterion with $\varepsilon_{{\rm stop},1} = 10^{-8}$ is used.

\subsection{Steering the Overlap to a Given Value}
\label{Subsection6.3}

\begin{figure}[ht!]
	\centering
	\includegraphics[width=1\linewidth]{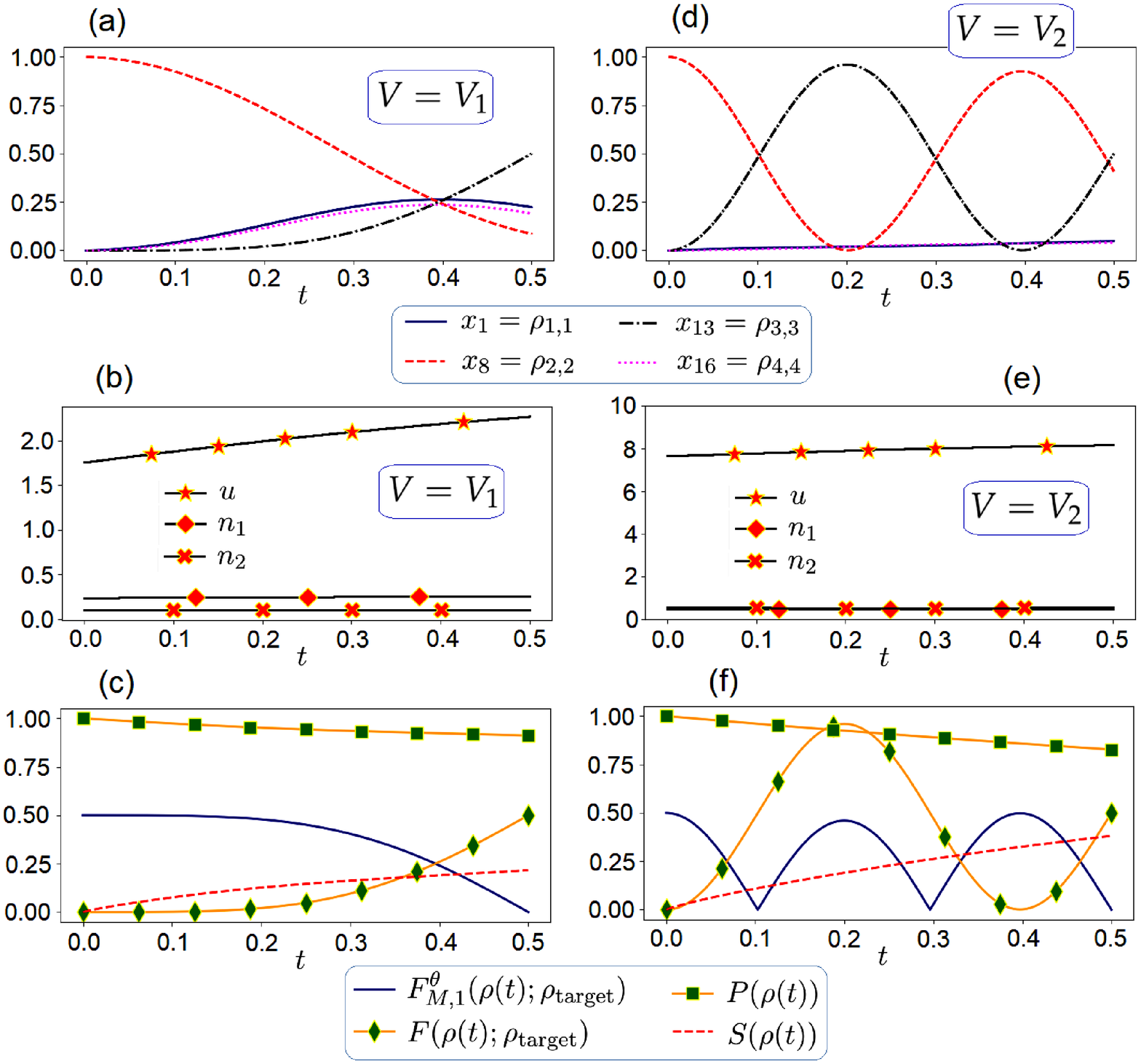} 
	\caption{The results of the GPM-2 for minimizing $J^{\theta}_{M,1}(c)$, where $\rho_0 = {\rm diag}(0,1,0,0)$, $\rho_{\rm target} = {\rm diag}(0, 0, 1, 0)$, $V$ is either $V_1$ or~$V_2$, and	$T=0.5$.}
	\label{Fig4}
\end{figure}

\begin{figure}[ht!]
	\centering
	\includegraphics[width=1\linewidth]{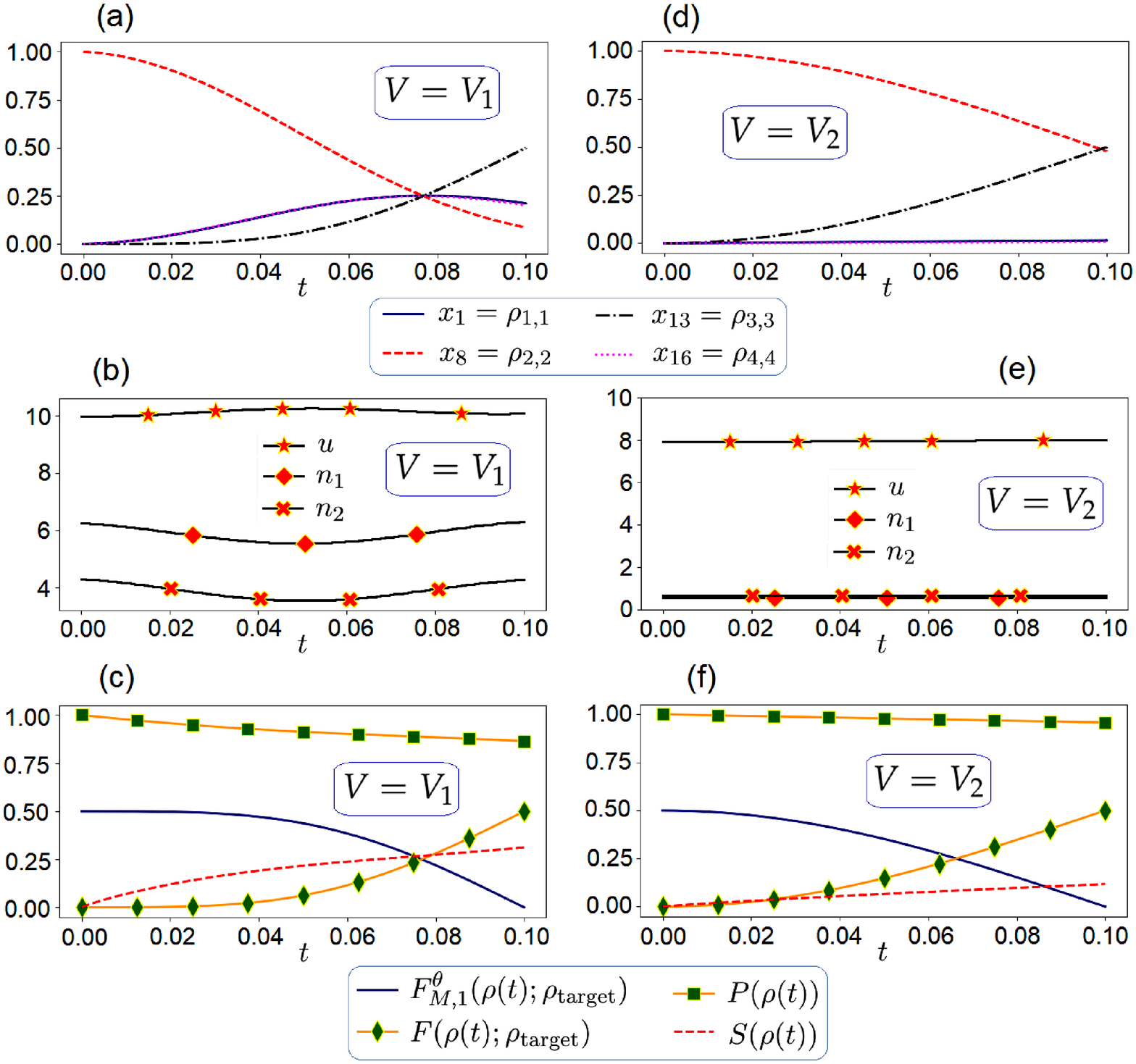} 
	\caption{The results of the GPM-2 for minimizing $J^{\theta}_{M,1}(c)$, where $\rho_0 = {\rm diag}(0,1,0,0)$, $\rho_{\rm target} = {\rm diag}(0, 0, 1, 0)$, $V$ is either $V_1$ or~$V_2$, and	$T=0.1$.}
	\label{Fig5}
\end{figure}

Consider the pure states $\rho_0 = {\rm diag}(0, 1, 0, 0)$ and $\rho_{\rm target} = {\rm diag}(0, 0, 1, 0)$, 
the operator $V$ being either $V_1$ or $V_2$, and $T$ being either 0.5~or~0.1, i.e. we have the four cases of $(V,T)$. For the problem of steering $\langle \rho(T), \rho_{\rm target} \rangle$ to $M = 0.5$, we use $I = J^{\theta}_{M,1}(c)$ with $\theta = 10^{-4}$. As (\ref{J_M_1_theta_inf}) shows, if $\vert\mathcal{F}(x(T); x_{\rm target}) - M\rvert \leq \theta$ we use the quadratic function majorizing values of $J_{M,1}(c)$. GPM-2 is used with $\alpha^{(k)} = \widehat{\alpha}/(k^{\sigma}+1)$, where $\sigma = 1.5$, $\beta = 0.92$ are for all the computations,  while $\widehat{\alpha}$ is specified for each of the four cases. Set $u^{(0)} = \sin t$, $n_1^{(0)} = n_2^{(0)} = 0$. For storing $u,~n_1,~n_2$ in computer, divide $[0,T]$ into $N = 500$ parts for $T=0.5$ and into $N = 100$ parts for $T = 0.1$, use piecewise constant interpolation. The values of $\widehat{\alpha}$ and the results of GPM-2 are shown in Table~\ref{Table2}, Fig.~\ref{Fig4},~\ref{Fig5}. We see that GPM-2 provides $J_{M,1}^{\theta}$ sufficiently close to~0 in each of the four cases. 

\begin{table}[ht!]
\begin{center}
\begin{minipage}{\textwidth}
\caption{About the work of GPM-2 for the problem of steering the overlap (Subsect.~\ref{Subsection6.3})}\label{Table2}
\begin{tabular*}{\textwidth}{@{\extracolsep{\fill}}lcccc@{\extracolsep{\fill}}}
\hline
 \backslashbox[0pt][l]{Data and Results}{Case $(V,T)$} & $(V_1, 0.5)$ & $(V_1, 0.1)$ & $(V_2, 0.5)$ & $(V_2, 0.1)$ \\\hline
     $I(c^{(0)})$ \--- initial value & $\approx 0.5$ & $\approx 0.5$ & $\approx 0.5$ & $\approx 0.5$ \vspace{0.2cm}\\   
    $\widehat{\alpha}$ in the formula for $\alpha^{(k)}$ & 1 & 100 & 1 & 5 \vspace{0.2cm}\\   
    $J_{M,1}(c)$ at the obtained $c$ & $\approx 6 \cdot 10^{-5}$ &  $\approx 5 \cdot 10^{-5}$ & $\approx 6 \cdot 10^{-5}$ & $\approx 6 \cdot 10^{-5}$ \vspace{0.2cm} \\  
    \begin{tabular}{@{}l@{}}Number of solving \\ the Cauchy  problems \end{tabular} & 169 & 243 & 345 & 275 \vspace{0.2cm}\\   
    $\aleph$ at the obtained $c$ & $\approx 0.21$ & $\approx 0.21$  & $\approx 0.11$  & $\approx 0.12$  \vspace{0.2cm} \\  
    Corresponding Fig. & \begin{tabular}{@{}c@{}}Fig.~\ref{Fig4}, \\ left col.\end{tabular} 
         & \begin{tabular}{@{}c@{}}Fig.~\ref{Fig5}, \\ left col.\end{tabular}
         & \begin{tabular}{@{}c@{}}Fig.~\ref{Fig4}, \\ right col.\end{tabular}
         & \begin{tabular}{@{}c@{}}Fig.~\ref{Fig5}, \\ right col.\end{tabular} \\ 
\hline
\end{tabular*}
\end{minipage}
\end{center}
\end{table}


For $V=V_1$, comparing the left columns in Figs.~\ref{Fig4},~\ref{Fig5}, we see that the resulting $\rho_{j,j}$ for the same $j \in \{1,2,3,4\}$ are similar to each other  for both variants of~$T$, and that for
the smaller~$T$ one has 
increasing in all the obtained controls' values. 
In Fig.~\ref{Fig4}, comparing the results
for $V = V_1$ (left column) with
the results for $V = V_2$ (right column) obtained for the same $T = 0.5$, we see the significant differences in
the obtained controls, etc. This was a reason for using $T = 0.1$ for $V = V_2$. The corresponding results are shown in the right column in Fig.~\ref{Fig5}, where the graphs of $\rho_{j,j}$ ($j=1,2,3,4$), $F_{M,1}^{\theta}(\rho(t); \rho_{\rm target})$, and $P(\rho(t))$ at the obtained $c$ are significantly more accurate. These observations remind the importance of comparing various numerically optimized control processes with respect to their structures. 

\section{Conclusions}
\label{Section7_Conclusions}

In this article, we consider optimal state manipulation for a two-qubit system whose dynamics is governed 
by the  Gorini--Kossakowski--Sudarshan--Lindblad master equation, where coherent 
control enters into the Hamiltonian and incoherent control into both the Hamiltonian (via Lamb shift) and the superoperator of dissipation. We exploit
two physically different classes of interaction with coherent control and consider 
the Hilbert--Schmidt overlap $\langle \rho(T), \rho_{\rm target} \rangle$ between final ($\rho(T)$) and target ($\rho_{\rm target}$) density matrices. We consider the problems of maximizing and minimizing the overlap and the problem of steering the overlap to a given admissible value. For the problems of optimizing the overlap, Theorem~\ref{Theorem1} contains the PMP, Theorem~\ref{Theorem2} contains conditions for $\rho_0$ and $\rho_{\rm target}$ such that zero coherent and incoherent controls satisfy the PMP, and Theorem~\ref{Theorem3} gives conditions for $\rho_0$ and $\rho_{\rm target}$  such that  zero controls form a stationary point of the objective functional. Known bounds for the overlap summarized in  Subsect.~\ref{subsection4.4}. For global optimization, these bounds can be useful for the analytical (Subsect.~\ref{Subsection4.6}) and numerical treatment. Sec.~\ref{Section5} describes one- and two-step GPM operating with piecewise continuous controls for the problems of maximizing the overlap and steering the overlap to a given value. The numerical results in Sec.~\ref{Section6} show the examples when the quantum system itself contains resources sufficient to reach the control goal without external controls (i.e., when zero coherent and incoherent controls are optimal), the gradient approach in the form of one- and two-step GPM is successful in our numerical experiments, coherent and incoherent controls can be simultaneously non-trivial, the profiles of the computed controls for various final times can behave essentially differently. 

\vspace{0.5cm}

\noindent {\bf Acknowledgements.} Sec.~\ref{Sec:IC} is performed in Steklov Mathematical Institute and supported by Russian Science Foundation under grant No.~22-11-00330,
\url{https://rscf.ru/en/project/22-11-00330/}. The other work was partially supported by the program in MIAN and the federal academic leadership program ``Priority 2030'' (MISIS Strategic Project Quantum Internet).


\begin{thebibliography}{99} 

\bibitem{KochEPJQuantumTechnol2022}
Koch, C.P., Boscain, U., Calarco, T., Dirr,~G., Filipp,~S.,
Glaser,~S.J., Kosloff,~R., Montangero,~S., Schulte-Herbr\"{u}ggen,~T.,
Sugny,~D., Wilhelm,~F.K.: Quantum optimal control in quantum technologies. 
Strategic report on current status, visions and goals for research 
in Europe. EPJ Quantum Technol. {\bf 9}, 19 (2022).
\url{https://doi.org/10.1140/epjqt/s40507-022-00138-x}

\bibitem{ButkovskiyBook1990} 
Butkovskiy, A.G., Samoilenko, Yu.I.: 
Control of Quantum-Mechanical Processes 
and Systems / Transl. from Russian. Kluwer Acad. Publ., 
Dordrecht (1990)

\bibitem{TannorBook2007} 
Tannor, D.J.: Introduction to Quantum Mechanics: 
A~Time Dependent Perspective. Univ. Science Books, Sausilito (2007).
\url{https://uscibooks.aip.org/books/introduction-to-quantum-mechanics-a-time-dependent-perspective/}

\bibitem{FradkovBook2007} 
Fradkov, A.L.: Cybernetical Physics: From Control of Chaos 
to Quantum Control. Springer, Berlin, Heidelberg (2007).
\url{https://doi.org/10.1007/978-3-540-46277-4}

\bibitem{WisemanBook2009} 
Wiseman, H.M., Milburn, G.J.: Quantum Measurement and Control.  
Cambridge Univ. Press, Cambridge (2009).
\url{https://doi.org/10.1017/CBO9780511813948}

\bibitem{BrifNewJPhys2010} 
Brif, C., Chakrabarti, R., Rabitz, H.:
Control of quantum phenomena: Past, 
present and  future. New J.~Phys. \textbf{12}(7), 075008 (2010).
\url{https://doi.org/10.1088/1367-2630/12/7/075008}

\bibitem{BonnardBook2012} 
Bonnard, B., Sugny, D.: Optimal Control with Applications in Space 
and Quantum Dynamics. AIMS, Springfield (2012).
\url{https://www.aimsciences.org/book/AM/volume/21}

\bibitem{KochJPhysCondensMatter2016}  
Koch, C.P.: Controlling open quantum systems: Tools, achievements, 
and limitations. J.~Phys.: Condens. Matter \textbf{28}(21), 213001 (2016).
\url{https://doi.org/10.1088/0953-8984/28/21/213001}

\bibitem{DAlessandroBook2021} 
D'Alessandro, D.: Introduction to Quantum Control and Dynamics. 
2nd Edition. Chapman and Hall/CRC, Boca Raton (2021). \url{https://doi.org/10.1201/9781003051268}

\bibitem{KuprovBook2023}
Kuprov, I.: Spin: From Basic Symmetries to Quantum Optimal Control.  
Springer, Cham (2023). \url{https://doi.org/10.1007/978-3-031-05607-9}

\bibitem{PalaoPRL2002}
Palao, J.P., Kosloff, R.: Quantum computing by an optimal control algorithm
for unitary transformations. Phys. Rev. Lett. {\bf 89}(18), 188301 (2002).
\url{https://doi.org/10.1103/PhysRevLett.89.188301}

\bibitem{TreutleinPRA2006}
Treutlein, P., H\"{a}nsch, T.W., Reichel,~J., Negretti,~A., Cirone,~M.A., Calarco,~T.: 
Microwave potentials and optimal control for robust quantum gates on an atom chip.  
Phys. Rev.~A {\bf 74}(2), 022312 (2006). \url{https://doi.org/10.1103/PhysRevA.74.022312}

\bibitem{Grace2007} 
Grace, M., Brif, C., Rabitz, H., Walmsley, I.A., Kosut, R.L., Lidar, D.A.: 
Optimal control of quantum gates and suppression of 
decoherence in a system of interacting two-level particles. 
J.~Phys.~B. {\bf 40}(9), S103 (2007). \url{https://doi.org/10.1088/0953-4075/40/9/S06}.

\bibitem{WuPRA2008}
Wu, R., Chakrabarti, R., Rabitz, H.: Optimal control theory for continuous-variable quantum gates.
Phys. Rev.~A {\bf 77}, 052303 (2008). \url{https://doi.org/10.1103/PhysRevA.77.052303}

\bibitem{PoulsenPRA2010}
Poulsen, U.V., Sklarz, S., Tannor,~D., Calarco,~T.:  
Correcting errors in a quantum gate with pushed ions via optimal control.
Phys. Rev.~A {\bf 82}(1), 012339 (2010). \url{https://doi.org/10.1103/PhysRevA.82.012339}

\bibitem{PechenJPA2017}
Pechen, A.N., Ilin, N.B.: Control landscape for ultrafast manipulation by a qubit. 
J.~Phys.~A {\bf 50}(7), 75301 (2017). \url{https://doi.org/10.1088/1751-8121/50/7/075301}

\bibitem{BasilewitschNewJPhys2019}
Basilewitsch, D., Cosco, F., Lo Gullo, N., M\"{o}tt\"{o}nen,~M.,
Ala-Nissil\"{a},~T., Koch,~C.P., Maniscalco,~S.:
Reservoir engineering using quantum optimal control for qubit reset.
New J. Phys. {\bf 21}, 093054 (2019). \url{https://doi.org/10.1088/1367-2630/ab41ad}

\bibitem{GoogleNature}
Arute, F., Arya, K., Babbush, R., et al.:
Quantum supremacy using a programmable superconducting processor.
Nature {\bf 574}, 505--510 (2019). \url{https://doi.org/10.1038/s41586-019-1666-5}

\bibitem{RiazQuantumInfProcess2019}
Riaz, B., Shuang, C., Qamar, S.: Optimal control methods for quantum 
gate preparation: a comparative study. Quantum Inf. Process. {\bf 18}, 100 (2019).
\url{https://doi.org/10.1007/s11128-019-2190-0}

\bibitem{VolkovJPA2021}  
Volkov, B.O., Morzhin, O.V., Pechen, A.N.: Quantum control landscape 
for ultrafast generation of single-qubit phase shift quantum gates.  
J.~Phys.~A {\bf 54}(21), 215303 (2021). \url{https://doi.org/10.1088/1751-8121/abf45d}

\bibitem{Volkov2022}
Volkov, B.O., Pechen, A.N.: On the detailed structure of quantum control 
landscape for fast single qubit phase-shift gate generation. 
Izv. RAN. Ser. Mat. (Accepted). \url{https://www.mathnet.ru/eng/im9372} 
(arXiv version: arXiv:2204.13671 [quant-ph] (2022). 
\url{https://doi.org/10.48550/arXiv.2204.13671})

\bibitem{LiScienceChina2013} 
Li, B., Yu, Z., Fei, S., Li-Jost, X.: Time optimal quantum 
control of two-qubit systems. Sci. China Phys. Mech. Astron.  
\textbf{56}, 2116--2121 (2013). \url{https://doi.org/10.1007/s11433-013-5325-9}

\bibitem{RafieePRA2016} 
Rafiee, M., Nourmandipour, A., Mancini,~S.:
Optimal feedback control of two-qubit entanglement 
in dissipative environments. Phys. Rev.~A \textbf{94}(1), 012310 (2016).
\url{https://doi.org/10.1103/PhysRevA.94.012310}

\bibitem{AllenPRA2017} 
Allen, J.L., Kosut, R., Joo, J., Leek,~P., Ginossar,~E.:
Optimal control of two qubits via a single cavity drive 
in circuit quantum electrodynamics. Phys. Rev.~A 
\textbf{95}(4), 042325 (2017). \url{https://doi.org/10.1103/PhysRevA.95.042325}

\bibitem{HuIJTP2018} 
Hu, J., Ke, Q., Ji, Y.: Steering quantum dynamics of a 
two-qubit system via optimal bang-bang control. 
Int.~J. Theor. Phys. \textbf{57}(5), 1486--1497 (2018).
\url{https://doi.org/10.1007/s10773-018-3676-8}

\bibitem{Hirose2018}
Hirose, M., Cappellaro, P.:
Time-optimal control with finite bandwidth.
Quantum Inf. Process. {\bf 17}, 88 (2018).
\url{https://doi.org/10.1007/s11128-018-1845-6}

\bibitem{FengPRA2018} 
Feng, G., Cho, F.H., Katiyar, H., Li,~J., 
Lu,~D., Baugh,~J., Laflamme,~R.:   
Gradient-based closed-loop quantum optimal control in
a solid-state two-qubit system. Phys. Rev.~A \textbf{98}(5), 052341 (2018).
\url{https://doi.org/10.1103/PhysRevA.98.052341}

\bibitem{SunQuantumInfProcess2020}
Sun, B.-Z., Fei, S.-M., Jing,~N., Li-Jost,~X.:
Time optimal control based on classification of quantum gates.
Quantum Inf. Process. {\bf 19}, 103 (2020).
\url{https://doi.org/10.1007/s11128-020-2602-1}

\bibitem{MorzhinLJM2021}   
Morzhin, O.V., Pechen, A.N.: Generation of density matrices 
for two qubits using coherent and incoherent controls. 
Lobachevskii J.~Math. \textbf{42}(10), 2401--2412 (2021).
\url{https://doi.org/10.1134/S1995080221100176}

\bibitem{ShindiIEEE2022} 
Shindi, O., Yu, Q., Dong, D.: A modified deep Q-learning 
algorithm for control of two-qubit systems. 
Proc. of 2021 IEEE Int. Conf. on Systems, Man, and Cybernetics 
(SMC), 3454--3459 (2021).
\url{https://doi.org/10.1109/SMC52423.2021.9658732}

\bibitem{PetruhanovIntJModPhysB2022}
Petruhanov, V.N., Pechen, A.N.: Optimal control for 
state preparation in two-qubit open quantum systems driven 
by coherent and incoherent controls via GRAPE approach. 
Int.~J. Mod. Phys.~B. \textbf{37}, 20n21, 2243017 (2022). 
\url{https://doi.org/10.1142/S0217751X22430175}

\bibitem{PRA84_042315}
M\"{u}ller, M.M., Reich, D.M., Murphy, M., Yuan,~H.,
Vala,~J., Whaley,~K.B., Calarco,~T., Koch,~C.P.:
Optimizing entangling quantum gates for physical systems.
Phys. Rev.~A {\bf 84}(4), 042315 (2011).
\url{https://doi.org/10.1103/PhysRevA.84.042315}

\bibitem{WangIndianJPhys2022} 
Wang, Q.L., Ren, H.F., Wang, P.:
Research on controlling the perfect transfer 
of the two-qubit quantum information in spin chain.
Indian J.~Phys. \textbf{96}(2), 391--397 (2022).
\url{https://doi.org/10.1007/s12648-020-01949-3}

\bibitem{SchirmerPRA2001} 
Schirmer, S.G., Fu, H., Solomon, A.I.: 
Complete controllability of quantum systems.
Phys. Rev. A {\bf 63}(6), 063410 (2001).
\url{https://doi.org/10.1103/PhysRevA.63.063410}

\bibitem{PolackPRA2009}
Polack, T., Suchowski, H., Tannor, D.J.: 
Uncontrollable quantum systems: A classification scheme based on 
Lie subalgebras. Phys. Rev.~A {\bf 79}(5), 053403 (2009).
\url{https://doi.org/10.1103/PhysRevA.79.053403}

\bibitem{KuznetsovLJM2022} 
Kuznetsov, S.A., Pechen, A.N.: On controllability of a highly 
degenerate four-level quantum system with a ``chained'' coupling 
Hamiltonian. Lobachevskii J. Math. {\bf 43}:7, 1683--1692 (2022).
\url{https://doi.org/10.1134/S1995080222100225}

\bibitem{MyachkovaFourLevel}
Myachkova, A.A., Pechen, A.N.: Some controllable and 
uncontrollable degenerate four-level quantum systems. 
Steklov Inst. Proc. (Accepted). 
\url{https://www.mathnet.ru/eng/tm4321} 

\bibitem{BochkinQIP2022} 
Bochkin, G.A., Fel'dman, E.B., Lazarev,~I.D., Pechen,~A.N.,
Zenchuk,~A.I.: Transfer of zero-order coherence matrix along 
spin-1/2 chain. Quantum Inf. Process. {\bf 21}, 261 (2022).
\url{https://doi.org/10.1007/s11128-022-03613-7}

\bibitem{Ai2014} 
Ai, Q., Fan, Y.-J., Jin, B.-Y., Cheng, Y.-C.: 
An efficient quantum jump method for coherent energy 
transfer dynamics in photosynthetic systems under 
the influence of laser fields. New J. Phys. \textbf{16}, 
053033 (2014). \url{https://doi.org/10.1088/1367-2630/16/5/053033}

\bibitem{BoscainPRXQuantum2021} 
Boscain, U., Sigalotti, M., Sugny,~D.: 
Introduction to the Pontryagin maximum principle 
for quantum optimal control. PRX Quantum \textbf{2}(3), 030203 (2021).
\url{https://doi.org/10.1103/PRXQuantum.2.030203}

\bibitem{BuldaevMathematics2022}   
Buldaev, A., Kazmin, I.: Operator methods of the maximum 
principle in problems of optimization of quantum 
systems. Mathematics \textbf{10}(3), 507 (2022).
\url{https://doi.org/10.3390/math10030507}

\bibitem{SchulteHerbruggenRevMathPhys2010}
Schulte-Herbr\"{u}ggen, T., Glaser,~S.J., Dirr,~G., Helmke,~U.: 
Gradient flows for optimisation in quantum information and 
quantum dynamics: Foundations and applications. 
Rev. Math. Phys. {\bf 22}(6), 597--667 (2010).
\url{https://doi.org/10.1142/S0129055X10004053}

\bibitem{Gough2005} 
Gough, J., Belavkin, V.P., Smolyanov, O.G.: 
Hamilton--Jacobi--Bellman equations for quantum optimal feedback control. 
J.~Opt.~B Quantum Semiclass. Opt. {\bf 7}(10), S237 (2005). 
\url{https://doi.org/10.1088/1464-4266/7/10/006}.

\bibitem{KrotovBook1996} 
Krotov, V.F.: Global Methods in Optimal Control Theory. 
New York, Marcel Dekker (1996)

\bibitem{MorzhinUMN2019}
Morzhin, O.V., Pechen, A.N.: Krotov method for optimal control 
of closed quantum systems. Russian Math. Surveys  
\textbf{74}(5), 851--908 (2019).
\url{https://doi.org/10.1070/RM9835}

\bibitem{Song2016} 
Song, X.-K., Ai, Q., Qiu, J., Deng, F.-G.: 
Physically feasible three-level transitionless quantum driving 
with multiple Schr\"{o}dinger dynamics. 
Phys. Rev.~A. \textbf{93}, 052324 (2016). 
\url{https://doi.org/10.1103/PhysRevA.93.052324}

\bibitem{Pechen_Prokhorenko_Wu_Rabitz_2008}
Pechen, A., Prokhorenko, D., Wu,~R., Rabitz,~H.:  
Control landscapes for two-level open quantum systems. 
J.~Phys.~A {\bf 41}, 045205 (2008). \url{https://doi.org/10.1088/1751-8113/41/4/045205}

\bibitem{Oza_Pechen_Dominy_Beltrani_Moore_Rabitz_2009}
Oza, A., Pechen, A., Dominy, J., Beltrani,~V., Moore,~K., Rabitz,~H.: 
Optimization search effort over the control landscapes for open quantum 
systems with Kraus-map evolution. J.~Phys.~A {\bf 42}, 205305 (2009).
\url{https://doi.org/10.1088/1751-8113/42/20/205305}

\bibitem{DannPRA2020}
Dann, R., Tobalina, A., Kosloff, R.:
Fast route to equilibration. Phys. Rev.~A \textbf{101}(5), 052102 (2020).
\url{https://doi.org/10.1103/PhysRevA.101.052102}

\bibitem{PechenPRA062102.2006} 
Pechen, A., Rabitz, H.: Teaching the environment to control 
quantum systems. Phys. Rev.~A \textbf{73}(6), 062102 (2006).
\url{https://doi.org/10.1103/PhysRevA.73.062102}

\bibitem{PechenPRA2011} 
Pechen, A.: Engineering arbitrary pure and 
mixed quantum states. Phys. Rev.~A \textbf{84}(4), 042106 (2011).
\url{https://doi.org/10.1103/PhysRevA.84.042106}

\bibitem{PechenPRA052102.2006} 
Pechen, A., Il'in, N., Shuang, F., Rabitz,~H.:
Quantum control by von Neumann measurements. 
Phys. Rev.~A \textbf{74}(5), 052102 (2006).
\url{https://doi.org/10.1103/PhysRevA.74.052102}

\bibitem{ShuangJChemPhys2007}  
Shuang, F., Pechen, A., Ho, T.-S., Rabitz,~H.: 
Observation-assisted optimal control of quantum dynamics. 
J.~Chem. Phys. \textbf{126}(13), 134303 (2007).
\url{https://doi.org/10.1063/1.2711806}

\bibitem{Shuang_Zhou_Pechen_Wu_Shir_Rabitz_2008}  
Shuang, F., Zhou, M., Pechen,~A., Wu,~R., Shir,~O.M., Rabitz,~H.: 
Control of quantum dynamics by optimized measurements. 
Phys. Rev.~A \textbf{78}(6), 063422 (2008).
\url{https://doi.org/10.1103/PhysRevA.78.063422}

\bibitem{HigginsNatComm2014}
Higgins, K.D.B., Benjamin, S.C., Stace,~T.M., Milburn,~G.J.,   
Lovett,~B.W., Gauger,~E.M.: Superabsorption of light 
via quantum engineering. Nat. Commun. {\bf 5}, 4705 (2014).
\url{https://doi.org/10.1038/ncomms5705}

\bibitem{HwangPRA2012}
Hwang, B., Goan, H.-S.: Optimal control for non-Markovian open 
quantum systems. Phys. Rev.~A {\bf 85}(3), 032321 (2012).
\url{https://doi.org/10.1103/PhysRevA.85.032321}

\bibitem{NJP17_063031}
Mukherjee, V., Giovannetti, V., Fazio,~R.,
Huelga,~S.F., Calarco,~T., Montangero,~S.:
Efficiency of quantum controlled non-Markovian thermalization.
New J. Phys. {\bf 17}, 063031 (2015).
\url{https://doi.org/10.1088/1367-2630/17/6/063031}

\bibitem{LucasRPL2013}
Lucas, F., Hornberger, K.: Adaptive resummation of Markovian 
quantum dynamics. Phys. Rev. Lett. {\bf 110}(24), 240401 (2013).
\url{https://doi.org/10.1103/PhysRevLett.110.240401}

\bibitem{LiningtonPRA2008}
Linington, I.E., Garraway, B.M.: Dissipation control in cavity 
QED with oscillating mode structures. Phys. Rev.~A {\bf 77}(3), 
033831 (2008). \url{https://doi.org/10.1103/PhysRevA.77.033831}

\bibitem{ZhongPRA2011}
Zhong, H., Hai, W., Lu, G., Li, Z.: 
Incoherent control in a non-Hermitian Bose-Hubbard dimer. 
Phys. Rev.~A {\bf 84}(1), 013410 (2011).
\url{https://doi.org/10.1103/PhysRevA.84.013410}

\bibitem{SinghPRA2007}
Singh, K.P., Rost, J.M.: Femtosecond photoionization of atoms 
under noise. Phys. Rev.~A {\bf 76}(6), 063403 (2007).
\url{https://doi.org/10.1103/PhysRevA.76.063403}

\bibitem{MukhopadhyayPRA2018}
Mukhopadhyay, C.: Generating steady quantum coherence and magic 
through an autonomous thermodynamic machine by utilizing a spin bath. 
Phys.~Rev.~A {\bf 98}(1), 012102 (2018).
\url{https://doi.org/10.1103/PhysRevA.98.012102}

\bibitem{JPA2007_8033}
Rossini, D., Calarco, T., Giovannetti,~V., Montangero,~S., Fazio,~R.:
Decoherence by engineered quantum baths.  
J.~Phys.~A {\bf 40}(28), 8033--8040 (2007).
\url{https://doi.org/10.1088/1751-8113/40/28/S12}

\bibitem{QutubuddinPRR2021}
Qutubuddin, Md., Dorfman, K.E.: Incoherent control of optical signals: 
Quantum-heat-engine approach. Phys. Rev. Res. 
{\bf 3}(2), 023029 (2021). \url{https://doi.org/10.1103/PhysRevResearch.3.023029}

\bibitem{Facchi2005} 
Facchi, P., Tasaki, S., Pascazio, S., Nakazato, H., Tokuse, A., Lidar, D.A.: 
Control of decoherence: Analysis and comparison of three different strategies. 
Phys. Rev.~A. {\bf 71}(2), 022302 (2005). 
\url{https://doi.org/10.1103/PhysRevA.71.022302}.

\bibitem{LaforgeJCP2018}
Laforge, F.O., Kirschner, M.S., Rabitz,~H.A.: 
Shaped incoherent light for control of kinetics: 
Optimization of up-conversion hues in phosphors. 
J.~Chem. Phys. {\bf 149}, 054201 (2018). \url{https://doi.org/10.1063/1.5035077}

\bibitem{Pechen_Trushechkin_2015}
Pechen, A.N., Trushechkin, A.S.: Measurement-assisted Landau--Zener 
transitions. Phys. Rev.~A {\bf 91}(5), 052316 (2015).
\url{https://doi.org/10.1103/PhysRevA.91.052316}

\bibitem{Mohseni2008} 
Mohseni, M., Rebentrost, P., Lloyd, S., Aspuru-Guzik, A. 
Environment-assisted quantum walks in photosynthetic energy transfer. 
J.~Chem. Phys. {\bf 129}(17), 174106 (2008). \url{https://doi.org/10.1063/1.3002335}.

\bibitem{LokutsievskiyJPA2021} 
Lokutsievskiy, L., Pechen, A.: Reachable sets for two-level 
open quantum systems driven by coherent and incoherent 
controls. J.~Phys.~A \textbf{54}(39), 395304 (2021).
\url{https://doi.org/10.1088/1751-8121/ac19f8}

\bibitem{MorzhinLJM2019} 
Morzhin, O.V., Pechen, A.N.: Maximization of the overlap between 
density matrices for a two-level open quantum system driven 
by coherent and incoherent controls. Lobachevskii J.~Math.
\textbf{40}(10), 1532--1548 (2019).
\url{https://doi.org/10.1134/S1995080219100202}

\bibitem{MorzhinPhysPartNucl2020} 
Morzhin, O.V., Pechen, A.N.: Maximization of the Uhlmann--Jozsa 
fidelity for an open two-level quantum system with coherent 
and incoherent controls. Phys. Part. Nucl. 
\textbf{51}(4), 464--469 (2020).
\url{https://doi.org/10.1134/S1063779620040516}

\bibitem{MorzhinIzvRAN2023}  
Morzhin, O.V., Pechen, A.N.: On optimization of coherent 
and incoherent controls for two-level quantum systems.  
Izv. RAN. Ser. Mat. (Accepted)  
(arXiv version: arXiv:2205.02521 [quant-ph] (2022).
\url{https://doi.org/10.48550/arXiv.2205.02521})

\bibitem{LevitinUSSRComputMathMathPhys1966}
Levitin, E.S., Polyak, B.T.: Constrained minimization methods. 
USSR Comput. Math. \& Math. Phys. \textbf{6}(5), 1--50 (1966).
\url{https://doi.org/10.1016/0041-5553(66)90114-5}

\bibitem{DemyanovBook1970} 
Demyanov, V.F., Rubinov, A.M.: Approximate Methods in Optimization Problems 
/ Transl. from Russian. American Elsevier Pub. Co., New York (1970)

\bibitem{PolyakBook1987} 
Polyak, B.T.: Introduction to Optimization / Transl. from Russian. 
Optimization Software Inc., Publ. Division, New York (1987)

\bibitem{PolyakUSSRComputMathMathPhys1964} 
Polyak, B.T.: Some methods of speeding up the convergence of 
iteration methods. USSR Comput. Math. \& Math. Phys.  
\textbf{4}(5), 1--17 (1964). \url{https://doi.org/10.1016/0041-5553(64)90137-5}

\bibitem{AntipinDifferEqu1994} 
Antipin, A.S.: Minimization of convex functions on convex sets 
by means of differential equations. Differ. Equat. 
\textbf{30}(9), 1365--1375 (1994)

\bibitem{HeutsIEEE2021} 
Heuts, Y.J.J., Padilla, G.P., Donkers,~M.C.F.: 
An adaptive restart heavy-ball projected primal-dual 
method for solving constrained linear quadratic optimal control problems. 
Proc. 60th IEEE Conf. CDC, 6722--6727 (2021).
\url{https://doi.org/10.1109/CDC45484.2021.9683013}

\bibitem{ZhangBook2011} 
Zhang, F.: Matrix Theory. Basic Results and Techniques.
2nd Ed. Springer, New York, Dordrecht, Heidelberg, London (2011).
\url{https://doi.org/10.1007/978-1-4614-1099-7}

\bibitem{NonsmootOpthBook2019}
Nonsmooth Optimization and Its Applications / 
Edt. by S.~Hosseini, B.S. Mordukhovich, A. Uschmajew.
Birkh\"{a}user (2019). \url{https://doi.org/10.1007/978-3-030-11370-4}

\bibitem{MorzhinAiT2009} 
Morzhin, O.V.: On approximation of the subdifferential 
of the nonsmooth penalty functional in the problems 
of optimal control. Autom. Remote Control \textbf{70}(5), 761--771 (2009).
\url{https://doi.org/10.1134/S0005117909050051}

\bibitem{MendoncaPRA2008} 
Mendon\c{c}a, P.E.M.F., Napolitano, R.d.J., Marchiolli,~M.A.,
Foster,~C.J., Liang,~Y.-C.: Alternative fidelity measure between
quantum states. Phys. Rev.~A \textbf{78}, 052330 (2008).
\url{https://doi.org/10.1103/PhysRevA.78.052330}

\bibitem{HolevoBook2019} 
Holevo, A.S.: Quantum Systems, Channels, Information: 
A Mathematical Introduction. 2nd Edition. De Gruyter, Berlin, Boston (2019).
\url{https://doi.org/10.1515/9783110642490}

\bibitem{WildeBook2017} 
Wilde, M.M.: Quantum Information Theory. 2nd Ed. 
Cambridge Univ. Press, Cambridge (2017).


\bibitem{PontryaginBook1962}   
Pontryagin, L.S., Boltyanskii, V.G., 
Gamkrelidze,~R.V., Mishchenko,~E.F.: 
The Mathematical Theory of Optimal Processes  / Transl. from Russian. 
Interscience Publishers John Wiley \& Sons, Inc., New York -- London (1962). 

\bibitem{GabasovBook1978}   
Gabasov, R., Kirillova, F.M.: Singular Optimal Controls / Transl. from Russian. 
Plenum Press, New York (1978)
\end{thebibliography}
\end{document}